\documentclass[3p]{elsarticle}

\usepackage{lineno,hyperref}

\usepackage{amsmath}
\usepackage{subfig}

\journal{NIM A}

\begin{document}

\begin{frontmatter}

\title{Benchmarking simulations of the Compton Spectrometer and Imager with calibrations}

\author[SSL]{Clio C. Sleator\corref{mycorrespondingauthor}}
\cortext[mycorrespondingauthor]{Corresponding author}
\ead{sleator@berkeley.edu}
\author[SSL,BIDS]{Andreas Zoglauer}
\author[UCSD]{Alexander W. Lowell}
\author[GSFC]{Carolyn A. Kierans}
\author[SSL]{Nicholas Pellegrini}
\author[SSL]{Jacqueline Beechert}
\author[UCSD]{Steven E. Boggs}
\author[GSFC]{Terri J. Brandt}
\author[SSL]{Hadar Lazar}
\author[UCSD]{Jarred M. Roberts}
\author[UCSD]{Thomas Siegert}
\author[SSL]{John A. Tomsick}

\address[SSL]{Space Sciences Laboratory, University of California, Berkeley, CA 94720-7450, USA}
\address[BIDS]{Berkeley Institute for Data Science, University of California, Berkeley, CA 94720-7450, USA}
\address[UCSD]{Center for Astrophysics and Space Sciences, University of California, San Diego, CA 92093, USA}
\address[GSFC]{NASA Goddard Space Flight Center, Greenbelt, MD 20771, USA}

\begin{abstract}
The Compton Spectrometer and Imager (COSI) is a balloon-borne $\gamma$-ray (0.2-5~MeV) telescope designed to study astrophysical sources. COSI employs a compact Compton telescope design utilizing 12 high-purity germanium double-sided strip detectors and is inherently sensitive to polarization. In 2016, COSI was launched from Wanaka, New Zealand and completed a successful 46-day flight on NASA's new Super Pressure Balloon. In order to perform imaging, spectral, and polarization analysis of the sources observed during the 2016 flight, we compute the detector response from well-benchmarked simulations. As required for accurate simulations of the instrument, we have built a comprehensive mass model of the instrument and developed a detailed detector effects engine which applies the intrinsic detector performance to Monte Carlo simulations. The simulated detector effects include energy, position, and timing resolution, thresholds, dead strips, charge sharing, charge loss, crosstalk, dead time, and detector trigger conditions. After including these effects, the simulations closely resemble the measurements, the standard analysis pipeline used for measurements can also be applied to the simulations, and the responses computed from the simulations are accurate. We have computed the systematic error that we must apply to measured fluxes at certain energies, which is 6.3\% on average. Here we describe the detector effects engine and the benchmarking tests performed with calibrations.
\end{abstract}


\end{frontmatter}


\section{Introduction}
Soft $\gamma$-rays (100~keV -- 10~MeV) are an excellent probe of the most extreme processes in our universe. With soft $\gamma$-rays, we can study non-thermal emission from neutron stars, Galactic black holes, and active galactic nuclei and advance our understanding of these exotic forms of matter. By studying soft $\gamma$-rays from radioactive decays within our Galaxy, we gain insight into fundamental physics and element formation in our Universe and how this ongoing nucleosynthesis is participating in the cycle of matter for next-generation stars \citep{Fryer19,Diehl13}. In addition, hot and otherwise untraceable phases of the interstellar medium can be studied. Despite the scientific richness of this regime of the electromagnetic spectrum, the challenges of observing astrophysical soft $\gamma$-rays are numerous. These include a high instrumental background, low interaction cross sections, and the need to conduct observations above the atmosphere \citep{BandstraThesis}. Due to these challenges, relatively few instruments have been designed to perform these intriguing observations.

\begin{figure}
\centering
\includegraphics[width=0.6\textwidth]{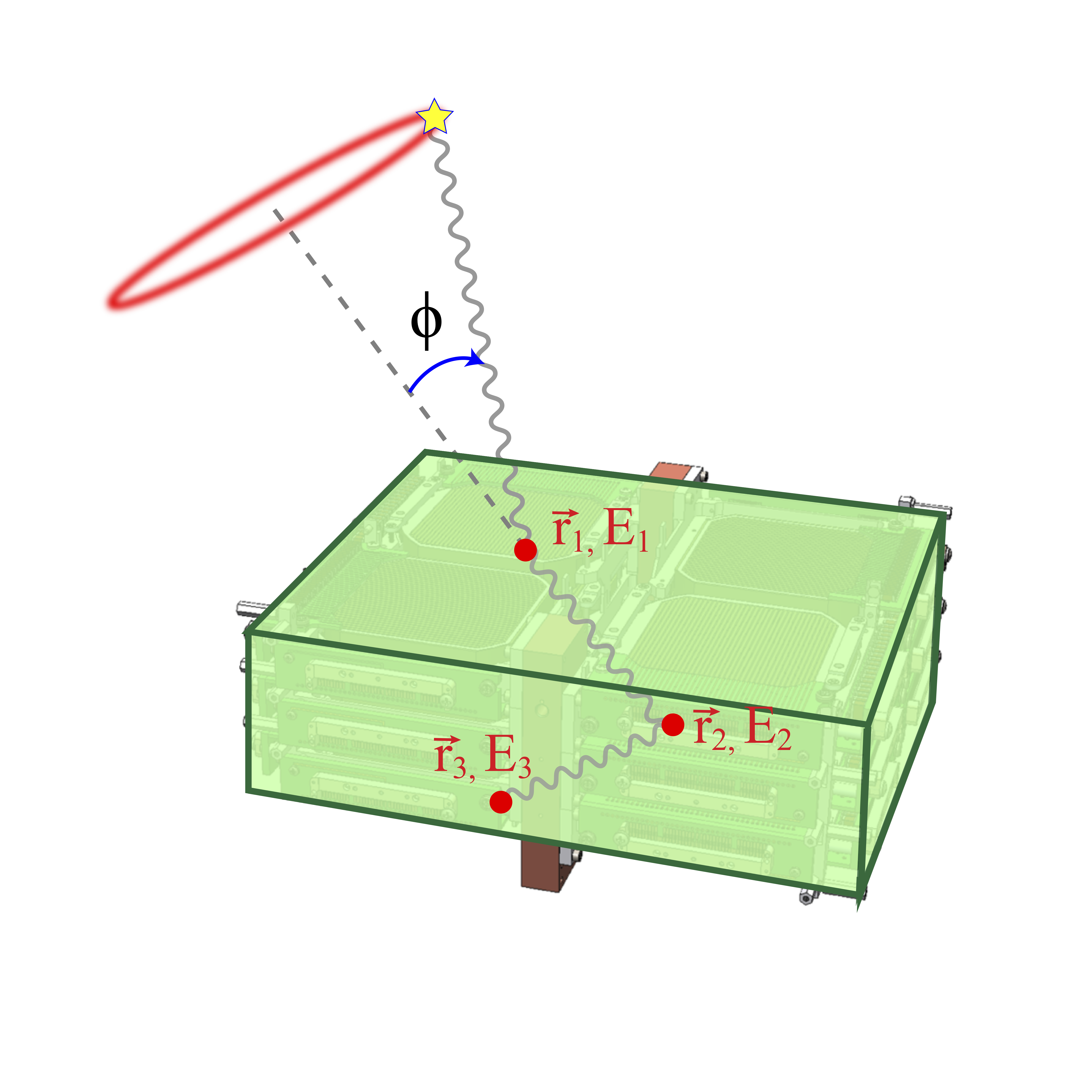}
\caption{A schematic of an ideal event in a compact Compton telescope. A photon originates from the star and Compton scatters twice in the detector, at $\vec{r}_1$ and $\vec{r}_2$, and deposits energy $E_1$ and $E_2$, respectively. The photon is then photoabsorbed at $\vec{r}_3$, depositing energy $E_3$. From the energy deposits at each interaction position, we can determine the initial Compton scatter angle $\phi$ and restrict the origin of the photon to the circle on the sky.}
\label{fig:cosiExp}
\end{figure}

Here we describe the Compton Spectrometer and Imager (COSI), a balloon-borne compact Compton telescope sensitive to $\gamma$-rays between 0.2 and 5~MeV \citep{LowellThesis,KieransThesis}. COSI's main science goals are to measure the polarization and spectrum of compact $\gamma$-ray sources such as $\gamma$-ray bursts, black holes, and neutron stars (e.g. \cite{McConnell17,Dean08,Laurent11}), to map the 511~keV positron annihilation line \citep{Siegert16,Kierans19}, to image diffuse emission from nuclear lines including $^{26}$Al and $^{60}$Fe \citep{Fryer19}, and to provide $\gamma$-ray sky coverage for multi-messenger astrophysics. Compton telescopes are a natural way to detect soft $\gamma$-rays, as these photons primarily interact with matter by Compton scattering. Compact Compton telescopes consist of one active detector volume in which a photon ideally Compton scatters at least once before being photoabsorbed \citep{BoggsJean00}. The order of interactions can be reconstructed using a variety of techniques, including kinematic reconstruction \citep{BoggsJean00}, Bayesian reconstruction \citep{ZoglauerThesis}, and machine learning \citep{Zoglauer07}. Once the interaction sequence is known, the origin of the photon can be constrained to the surface of a cone using the classic Compton scattering formula, which can then be projected into a circle on the sky (see Figure \ref{fig:cosiExp}). Additional tracking of the recoil electron would reduce the circle to an arc on the sky, which is beyond COSI's current design. The location of a source emitting multiple photons is at the point of overlap of the resulting Compton circles. Iterative deconvolution techniques can be employed to determine the most likely source distribution \citep{Zoglauer11}. As a Compton telescope, COSI is inherently sensitive to polarization, as the distribution of azimuthal scattering angles in the detector is dependent on the photon's polarization direction \citep{Lei97}.

On May 17, 2016, COSI was launched from Wanaka, New Zealand on NASA's Super Pressure Balloon and had a successful 46-day flight \citep{Kierans16}. The astrophysical sources detected during the 2016 flight include the positron annihilation emission from the 511~keV line as well as the low-energy ortho-positronium decay continuum \citep{KieransThesis}, a long, bright GRB \citep{Tomsick16,Lowell17a}, and a few compact objects, including the Crab Nebula. To perform spectral, polarization, and imaging analysis of these sources and of any additional sources detected during future COSI flights, well-benchmarked simulations with which we compute the instrument response are essential. Additionally, we use these simulations to benchmark and improve the data analysis pipeline and to better understand the instrument calibrations and in-flight performance.

As required for the simulations, we have built a comprehensive mass model of the instrument and developed a detailed detector effects engine (DEE) which applies the intrinsic detector performance (e.g. finite energy, timing, and position resolution) to Monte Carlo simulations. With the addition of the DEE, the simulations closely resemble the measurements and thus can be used to compute an accurate instrument response.  In a balloon-borne experiment such as COSI, our target accuracy between simulations and measurements is a systematic error of 10\% or less. This paper describes the simulation pipeline and the DEE, and presents the benchmarking tests that we performed to ensure that the simulations closely match the calibration data.

\section{The COSI instrument}
\begin{figure}
\centering
\includegraphics[width=0.35\textwidth]{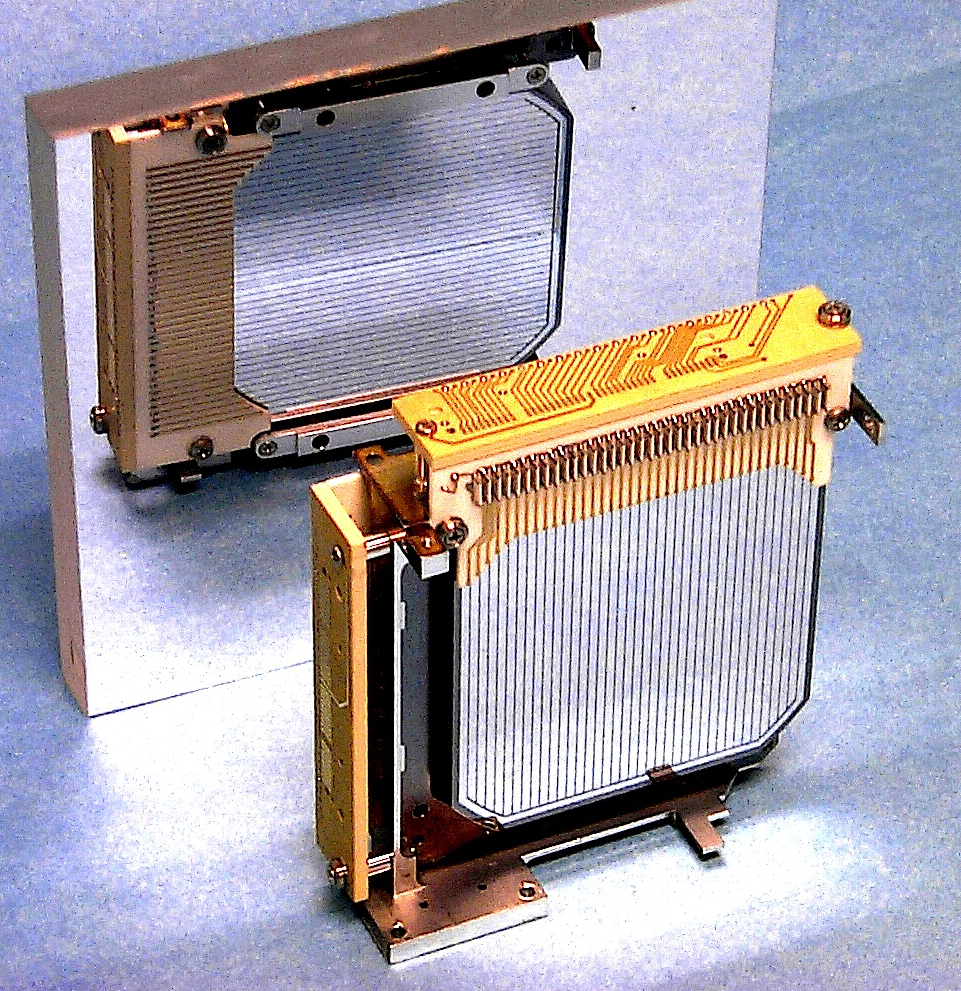}
\caption{A single COSI GeD in front of a mirror, showing the orthogonal cross strips.}
\label{fig:ged}
\end{figure}

COSI's active detector volume consists of a $2\times2\times3$ array of 12 high-purity germanium detectors (GeDs), each with a volume of $8\times8\times1.5$~cm$^3$ (see Figure \ref{fig:ged}). The anode and cathode electrodes on each side of the detector are segmented into 37 strips with a strip pitch of 2 mm. The strips on the anode are deposited orthogonally to those on the cathode so that the $x$ and $y$ interaction position can be determined using the positions of the triggered strips. A 2~mm guard ring surrounds each side of the detector to prevent leakage current from flowing between the anode and cathode. A high voltage between 1000 and 1500~V is applied to each GeD. On the cathode side of the detector, the charge sensitive preamplifiers are AC coupled to the detector strips in order to block the high voltage bias, meaning that a capacitor filters out the low frequency and DC offset components; we refer to this side as the AC side. On the anode, the charge sensitive preamplifiers are DC coupled to the detector strips, allowing the entire signal to pass through, and so we refer to the anode as the DC side.

The detectors are housed in an aluminum cryostat and are kept at cryogenic temperatures ($\sim$84 K) as required for GeDs with a mechanical cryocooler, enabling ultra-long duration balloon flights. The cryostat bottom and sides are enclosed by a cesium iodide (CsI) active anti-coincidence shield system to reduce atmospheric background. The CsI shields limit COSI's field of view to 25\,\% of the sky.

Each strip on each side of each detector ($37\times2\times12=888$ strips total) is instrumented with a charge sensitive preamplifier followed by two shapers: a fast bipolar shaper that measures the relative collection time of the charge carriers, and a slow unipolar shaper that precisely measures the pulse height of the signal with minimal noise. The pulse shaping analog electronics and the trigger logic for a single detector are housed in a ``card cage", which sends the event data to the flight computer. For the event to proceed from the card cage to the flight computer, at least one strip on each side of the detector must trigger the fast shapers within a 360~ns window. The size of the trigger window was selected empirically to allow charge carriers from all events to reach the electrodes, regardless of where in the detector the event occurred. After this 360~ns window, the card cage waits 2.4~$\mu$s for veto signals from the anti-coincidence shields or the detector guard ring. The veto windows were tuned by measuring the time delay between the veto signal and card cage trigger signal, and then set to encompass the maximum time delay. If no veto signals occur during the veto windows, the pulse heights are accurately measured with the slow shapers and the event information is sent from the card cage to the flight computer.

\section{Data analysis pipeline}
\label{sec:calib}
\begin{figure}
\centering
\includegraphics[width=\textwidth]{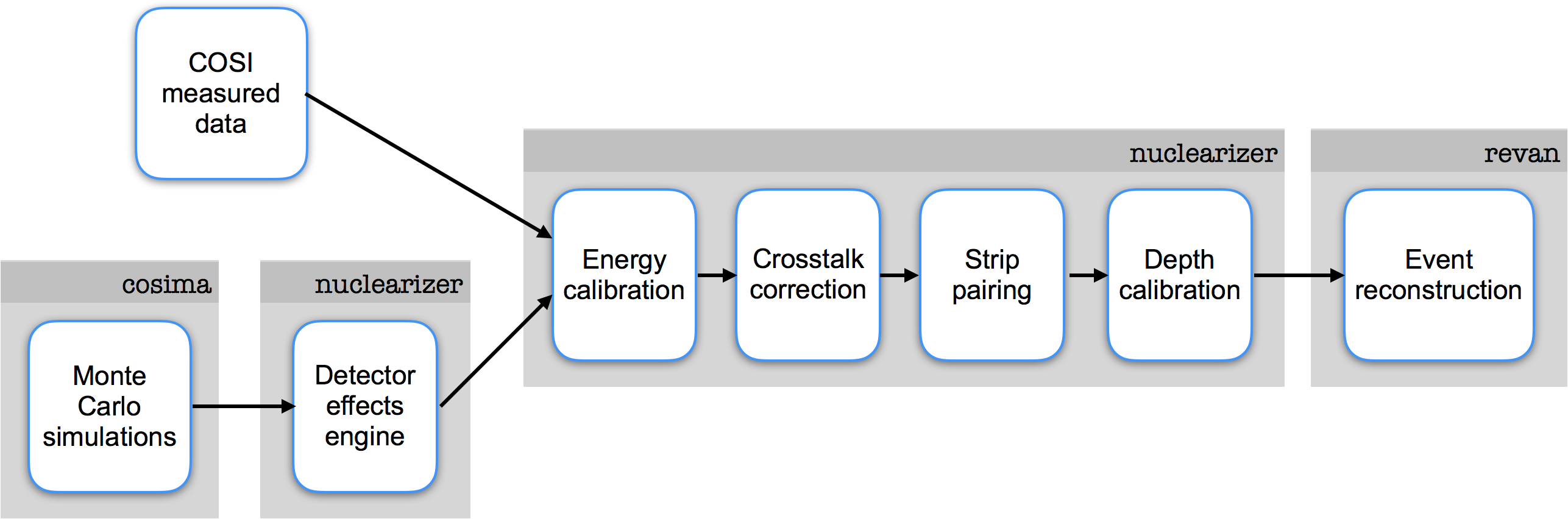}
\caption{The analysis pipeline for COSI data. Simulated data goes through the detector effects engine such that the simulations resemble measured data. Both the simulations and measured data go through the event calibration and event reconstruction steps of the pipeline. After the event reconstruction, the data can be used for high level analysis such as spectral or polarization analysis.}
\label{fig:analysisPipeline}
\end{figure}

The COSI analysis tools are built on the Medium Energy Gamma-ray Astronomy Library (MEGAlib) \cite{MEGAlib} software specifically designed for the analysis of data from Compton telescopes. Figure \ref{fig:analysisPipeline} shows the data analysis pipeline used for COSI. Both measurements and simulations go through the same event calibration and event reconstruction steps.

The event calibration program \emph{nuclearizer} converts the measured parameters of pulse height, pulse timing, detector ID, and strip ID into the physical parameters of energy deposited and position within the detector. The event calibration steps are:

\begin{enumerate}
\item \textbf{Energy calibration:} The pulse height is proportional to the energy deposited in the interaction, so we can determine the pulse height-energy conversion using calibration sources that emit $\gamma$-rays at known line energies. We fit the pulse height-energy relation for each strip with an empirical model that accounts for any small non-linear deviations \citep{Kierans14}.
\item \textbf{Crosstalk correction:} Crosstalk is the influence of one electronics channel on another. In the case of the COSI detectors, if two nearby strips trigger, the energy recorded is amplified for both strips and the total energy is higher than it would be if all the energy was deposited on a single strip. Since the crosstalk effect is linear with energy, correcting it is straightforward, as described in \cite{BandstraThesis}.
\item \textbf{Strip pairing:} Strip pairing determines the $x$-$y$ interaction position from the AC and DC strip IDs. If there is only one interaction in a single detector, this process is straightforward: the interaction occurred where the two strips intersect. If a photon interacts multiple times in the detector, determining the interaction locations can be complicated. In general, we pair the strips by comparing their energies, as each interaction should result in an equal amount of charge deposited on the AC and DC strips.
\item \textbf{Depth calibration:} The depth calibration determines the $z$ position, or depth in the detector, as a function of the collection time difference (CTD), which is the difference in collection times of the electrons on the anode and the holes on the cathode. Calibrating the CTD-depth relation is described in \cite{Lowell16}.
\end{enumerate}

The event reconstruction is done using \emph{revan}, the real event analyzer for MEGAlib. \emph{Revan} determines the order of interactions within the detector volume \citep{BoggsJean00,ZoglauerThesis,Zoglauer07} and thus the initial Compton scatter angle.

The simulation pipeline consists of three main steps before the event calibration and event reconstruction: the mass model, Monte Carlo simulations, and the DEE. The mass model dictates that the correct amount of material is at the correct position and in the correct shape, and thus determines where in the detectors the simulated interactions occur. Because $\gamma$-rays interact with passive material as well as with active detectors, it is important to model all objects near the detectors, including but not limited to the cryostat shell, cryocooler, readout electronics, gondola frame, and gondola subsystems. To validate the mass model, we ensure that simulations of calibration sources accurately reproduce the measurements regardless of their position relative to the instrument, as $\gamma$-rays emitted by off-axis sources go through more and different material than $\gamma$-rays emitted by a source directly on-axis. The results of this validation are described in Section \ref{sec:simulationBenchmarking}.

Monte Carlo simulations of calibration runs and astrophysical observations are performed using \emph{cosima}, the Cosmic Simulator for MEGAlib based on Geant4 \citep{Geant4}. \emph{Cosima} performs Monte Carlo simulations of various source spectra and geometries and can simulate space, balloon, and lab environments. As input, \emph{cosima} requires the mass model, the source position, which can be an astrophysical or local position (i.e. far field or near field), and the source emission properties, including the energy spectrum, flux, polarization, and distribution of emission directions. The \emph{cosima} output is an event list describing interactions in the detectors as defined by the mass model.

MEGAlib includes two ways of processing the \emph{cosima} output: a perfect reconstruction and the standard MEGAlib DEE. The perfect reconstruction takes the event list and performs the event reconstruction without adding any noise, e.g. energy resolution. The standard MEGAlib DEE applies the average energy resolution, depth resolution, and thresholds per detector to the simulations, but does not apply any other specific detector and readout electronics effects. We have developed a COSI-specific DEE that includes many of these other effects, including charge sharing, charge loss, crosstalk, and dead time. The COSI-specific DEE also uses the measured energy resolution and threshold values of each individual strip rather than an average value over the entire detector. Lastly, the COSI-specific DEE reverses the event calibration by converting the physical parameters of energy and position into the measured parameters of pulse height, timing, strip ID, and detector ID. After this conversion, the simulation format mimics the data format and the simulations are run through the COSI event calibration pipeline. Thus, any imperfections present in the calibration pipeline will affect the simulations as well as the measurements. With the COSI-specific DEE, described in detail in the next section, the simulations much more accurately reproduce the measurements.

We note that accurate simulations are one key step required to analyze astrophysical data measured by COSI, and that to perform spectral and polarization analysis of astrophysical sources, we must also employ a careful treatment of the background that is so prevalent in the MeV regime. The details of this background treatment are beyond the scope of this work, but are discussed in e.g. \cite{KieransThesis}. Throughout the remainder of this work, we benchmark the simulations using calibration sources that emit $\gamma$-rays, and thus, for example, ignore charged particle events. However, the vast majority of charged particle events are rejected during the analysis since they either interact too many times in the detector or are vetoed by the shields. Secondaries such as Bremsstrahlung are typically not be rejected, but these are normal photon interactions that create signatures similar to the calibration sources.

\section{The COSI detector effects engine}
\label{sec:dee}

Here we describe the steps of the COSI-specific DEE, referred to simply as the DEE in this section, in the order in which they occur in the code. We adopt the following terminology: an \emph{event} is a single photon that interacts in the detectors. Each interaction is referred to as a \emph{hit}, and an event can have one hit (if the $\gamma$-ray is immediately photoabsorbed) or multiple hits (if the $\gamma$-ray Compton scatters in the detector before being photoabsorbed). Each hit contains multiple \emph{strip hits}, which refer to the individual strips that trigger during an interaction. A hit must contain at least one strip hit per detector side, but can also contain multiple strip hits on each detector side in the case of charge sharing between neighboring strips.

Throughout this section, we compare simulations to calibration data taken in the lab. To take calibration data, we use Type D disk $\gamma$-ray sources from Eckert \& Ziegler that emit $\gamma$-rays at known line energies (between 60~keV and 1.836~MeV) and with well-calibrated activities, and place them in a variety of locations in COSI's field of view.

Accurately modeling the calibration sources in the mass model is essential for comparisons of this sort, as $\gamma$-rays leaving the source could interact with nearby passive material. Initial scatters off of this material could, for example, change the observed flux of X-ray lines, such as the 32.3~keV line from $^{137}$Cs. We include the evaporated metallic salt material and the plastic casing of the calibration sources in the mass model. The sources are held in place by a metal structure made out of 80/20 T-slot material and attached to the structure with a piece of plastic. The plastic source holder piece is the closest part of the structure to the source, so $\gamma$-rays leaving the source could scatter off of it or be absorbed in it; thus we include it in the mass model. We also considered the relationships of the emitted $\gamma$-ray lines from a single source. For example, in the case of $^{22}$Na, although the branching ratio of the 1275~keV line from Na-22 is 100\%, 90\% of these decays are effectively coincident with a $\beta^+$ decay, which results in two diametrically emitted 511~keV $\gamma$-rays. This process, and similar processes for other calibration sources, were included in the Monte Carlo simulations generated by \emph{cosima}.

\subsection{Shield veto}
If there is a coincident interaction in the CsI shields and in the GeDs, the event is vetoed, as it is not possible to properly reconstruct events that do not deposit all of their energy in the GeDs. After each shield event, an analog shield veto signal remains active for a certain amount of time, dependent on the energy. After each GeD event, the card cages wait for 0.4~$\mu$s to receive the shield veto signal. If the shield veto signal is active at any point during this 0.4~$\mu$s coincidence window, the GeD event is vetoed. This process is simulated in the DEE, where the shield veto signal remains active for 1.7~$\mu$s after each shield event. This 1.7~$\mu$s duration is the measured average shield dead time per event.

\subsection{Physical position to detector and strip ID}
The event list that \emph{cosima} outputs provides a physical $(x, y, z)$ position for each hit. The physical position is converted into the corresponding detector ID and the closest AC and DC strip ID. Each of the two resulting strip hits is assigned the energy deposited in this interaction. We also calculate the depth in the detector for this interaction: based on the $z$ position of interaction within the instrument, we determine the depth within the activated detector based on its relative position in the mass model. From the depth in the detector, we determine the relative timing of the simulated strip hits (see Section \ref{sec:depthToTiming}) so that the simulated format mimics the format of the measured data. The $z$ position in the detector is then reestablished during the event calibration.

\subsection{Charge sharing}
\label{sec:chargesharing}
Charge sharing describes the phenomenon of a single interaction triggering multiple adjacent strips. This effect is attributed to several physical processes. Thermal diffusion and charge carrier repulsion can cause the charge cloud to spread laterally \citep{Knoll}. Amman and Luke (2000) \cite{Amman00} measured charge sharing in cross strip GeDs and attributed the effect to interactions that physically occur between strips. Work by Looker et al. (2015) \cite{Looker15} indicates that different detector fabrication techniques can increase or mitigate charge sharing.

Including charge sharing in the DEE is important for a number of reasons. It is difficult to properly reconstruct events that contain charge sharing hits, where charge is collected over two or more adjacent strips. Charge loss and crosstalk are effects that occur in the COSI GeDs and distort the measured energy of adjacent strips. If the measured energy of the hit on one side of the detector is significantly higher or lower than the measured energy of the hit on the other side, the strip pairing calibration (see Section \ref{sec:calib}) is not able to pair the strips properly and flags the event; these flagged events are discarded later in the analysis pipeline. Additionally, low energy hits may deposit an energy above the strip threshold if all the charge is collected on one strip but below the threshold if the energy is split into two strips. Thus, charge sharing can lead to sub-threshold hits which are not measured by the GeDs, mimicking incompletely absorbed events that are difficult to reconstruct.

Simulating the majority of the effects that cause charge sharing is difficult without a detailed charge transport simulation that takes into account variations in the electric field, charge carrier repulsion, and any effects caused by the detector fabrication method. It may be possible to characterize the charge sharing effect empirically by illuminating each GeD with a collimated beam, as done in \cite{Amman00}, but this experiment was not performed before the 2016 flight. Simulating charge sharing due to thermal diffusion, however, is relatively straightforward. Thermal diffusion introduces some spread in the arrival position of the charge carriers. The spread can be characterized as a Gaussian with a width of

\begin{equation}
\label{eq:diffusion}
\sigma_\text{diffusion}=\sqrt{\frac{2kTz}{eE}}
\end{equation}

\noindent where $k$ is Boltzmann's constant, $T$ is the detector temperature, $z$ is the distance that the charge carrier travels along the electric field direction (the $\hat{z}$ direction), $e$ is the electron charge, and $E$ is the electric field applied to the detector \citep{Knoll}. To simulate this effect, we divide the deposited energy into charge carriers. For each charge carrier, we randomly draw an $x$-$y$ drift position from a 2-dimensional Gaussian distribution with width $\sigma_\text{diffusion}$ and centered at the $x$-$y$ interaction position. We then determine the AC strip and DC strip that correspond to the $x$-$y$ drift position, and add the energy of the charge carrier to those strips.

The effects of charge sharing due to diffusion, however, are small in the COSI GeDs. For a detector temperature $T=84$~K, an electric field $E=1000$~V/cm, and an interaction occurring in the middle of the detector at $z=0.75$`cm, the resulting Gaussian has a width of $\sigma_\text{diffusion}=0.033$~mm. With a strip pitch of 2~mm, charge sharing due to diffusion is unlikely to have a significant effect. Nevertheless, this method can be used to empirically simulate all of the charge sharing contributions in the detectors by increasing  $\sigma$ by a factor $N$:

\begin{equation}
\label{eq:empiricalChargeSharing}
\sigma=N\sigma_\text{diffusion}=N\sqrt{\frac{2kTz}{eE}}~~.
\end{equation}

\noindent The scaling factor $N$ depends on the detector and deposited energy. For each detector, $N$ was empirically selected at each calibration source line energy so that the number of adjacent strip hits in the simulation is equal to that in the data. To find $N$ at any energy, we linearly interpolate between the values at the calibration source energies. The value of $N$ averaged over the 12 GeDs ranges from 3.28 at 122~keV to 4.02 at 1.2~MeV on the AC side, and from 1.70 at 122~keV to 1.95 at 1.2~MeV on the DC side.

\begin{figure}
\centering
\subfloat[]{\includegraphics[width=0.5\textwidth]{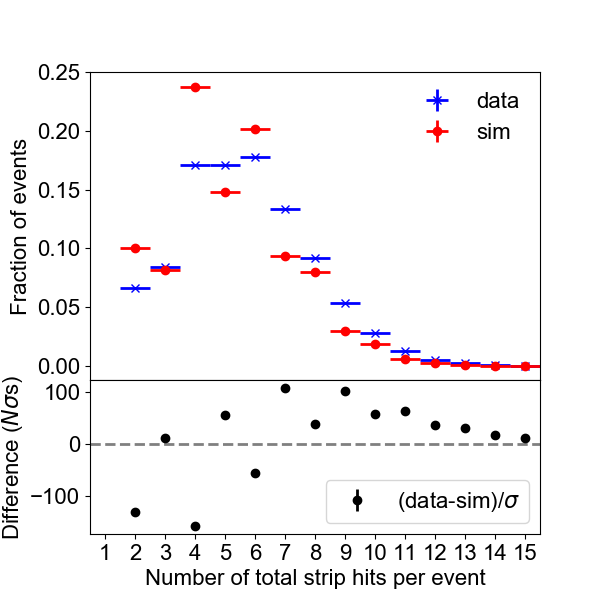}}
\subfloat[]{\includegraphics[width=0.5\textwidth]{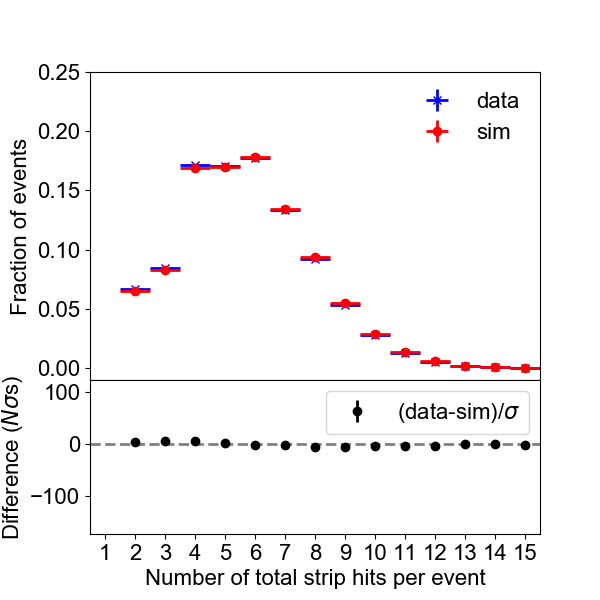}}
\caption{Measured and simulated distribution of the number of strip hits per event (a) without charge sharing and (b) with charge sharing included in the simulations. The included events are from the 662 keV line of a $^{137}$Cs source. Charge sharing was empirically applied to the simulations assuming that the charge carriers spread according to a Gaussian distribution with a width as described in Equation \ref{eq:empiricalChargeSharing}, and there is very good agreement between measurements and simulations.}
\label{fig:diffusionFactor}
\end{figure}

Figure \ref{fig:diffusionFactor} compares the measured and simulated distributions of the number of strip hits per event from the 662~keV line of a $^{137}$Cs source before and after charge sharing has been included in the simulations. Charge sharing affects this distribution, as it will cause more events to have a larger number of strip hits per event. Note that a single event often contains multiple hits and each hit contains at least two strip hits; thus one event can contain many strip hits. After charge sharing is added to the simulations, these measured and simulated distributions match well and residuals are significantly smaller, indicating that this empirical method of simulating charge sharing is a good approximation to the physical charge sharing in the COSI GeDs.

\begin{figure}
\centering
\includegraphics[width=0.5\textwidth]{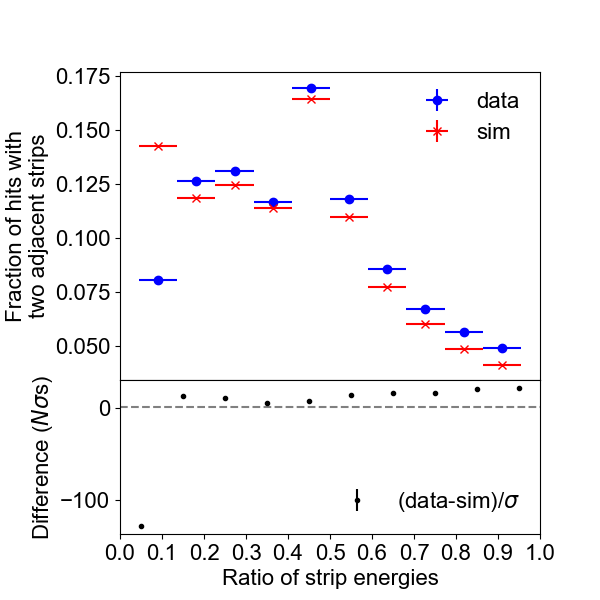}
\caption{Measured and simulated distribution of the energy between two adjacent strips for hits containing two adjacent strips. The energy ratio is the energy of the strip hit with lower energy divided by the energy of the strip hit with higher energy. The agreement between measurements and simulations would likely be improved by a more physical model of charge sharing, particularly the the agreement at a low energy ratio, which corresponds to a large difference in energy between the adjacent strip hits.}
\label{fig:energyRatio}
\end{figure}

Figure \ref{fig:energyRatio} shows the ratio in energies between the two adjacent strip hits of both measured and simulated hits that contain exactly two adjacent strip hits. The histograms are scaled by the total number of hits with two adjacent strip hits. There is some discrepancy between the measured and simulated distributions of energy ratios, especially at low ratios, which correspond to a large difference in energies between the two strip hits. This discrepancy likely occurs because we used a Gaussian distribution to simulate charge sharing rather than a more physical distribution. Although a more physical simulation of charge sharing would like reduce these discrepancies between measurements and simulations, determining such a distribution requires a detailed charge transport simulation of the COSI GeDs, which is beyond the scope of this work.

\subsection{Depth to timing}
\label{sec:depthToTiming}
We invert the depth calibration (see Section \ref{sec:calib}) to convert depth into CTD, or the difference between the charge collection times on the AC and DC strips. Though only the CTD is used in the event calibration, each strip hit must be assigned a timing value such that the output format of the DEE accurately mimics the real data format. Because the timing shaper of each strip has a unique offset that is not individually calibrated, the timing value of each strip hit is not meaningful. (Note that the event time is determined independently of the strip hit time: each card cage receives a 10~MHz clock signal from the flight computer which is used to determine the event time). Thus, we assign each strip an arbitrary timing while ensuring the correct CTD. We then apply Gaussian noise to the timing by randomly drawing a number from a Gaussian distribution centered at zero with a width of 12.5~ns and add that randomly drawn number to the strip hit timing value. The 12.5~ns Gaussian width is determined by the depth calibration \citep{BandstraThesis}.

\begin{figure}
\centering
\subfloat[]{\includegraphics[width=0.5\textwidth]{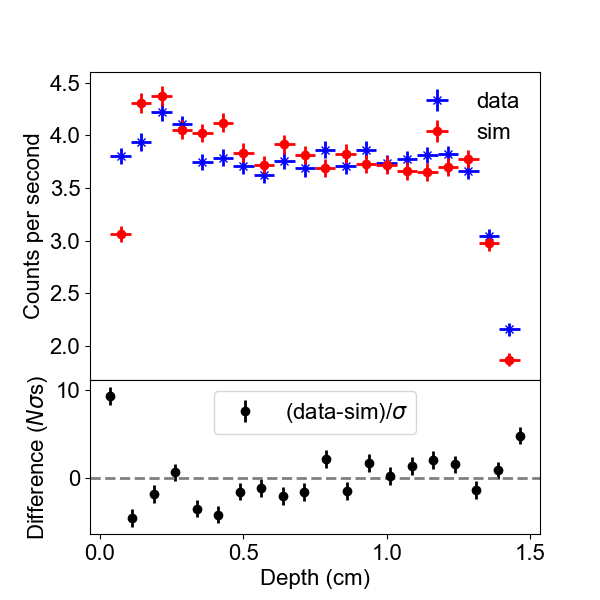}}
\subfloat[]{\includegraphics[width=0.5\textwidth]{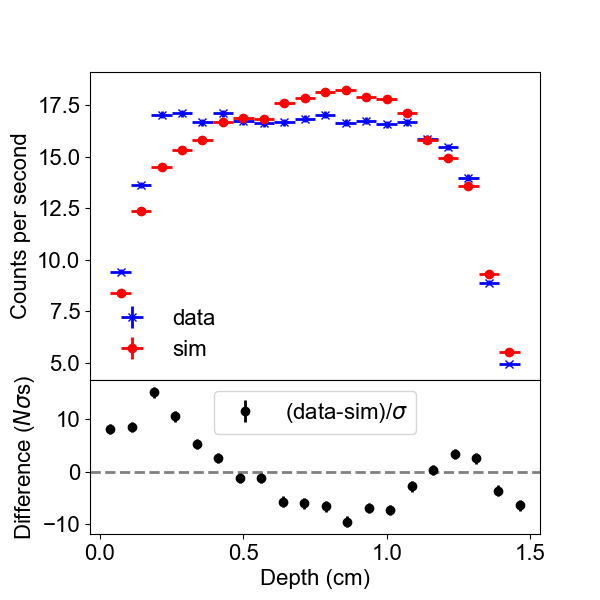}}
\caption{Measured and simulated depth distributions for the top layer of detectors (a) only including hits with one strip hit per side and (b) including all hits. The distributions for the middle and bottom detector layers have the same shape.}
\label{fig:depthDist}
\end{figure}

Figure \ref{fig:depthDist} compares the measured and simulated depth distributions after the depth calibration step. The distributions match well for hits with only one strip hit per side. When all hits are included, however, the differences between the measured and simulated distributions increase. Improvements to the depth calibration, which are beyond the scope of this work, could potentially alleviate this issue. The CTD-depth relationship is determined with hits that contain only one strip hit per side, meaning that we do not have a precise depth calibration for charge sharing hits. It is likely that applying the CTD-depth relationship to charge sharing hits skews the measured depth distribution, especially as initial investigations have shown that the timing measurement on a single strip changes when neighboring strips also collect charge.

\subsection{Charge loss}
Charge loss occurs when not all of the charge deposited in the detector crystal is read out by the electrodes. This effect occurs for a number of reasons. The charge carriers from interactions that occur in between two strips may not arrive at the electrodes within the charge collection time, due to the lack of an electric field in the plane of the strips \citep{Amman00}. Phenomena that cause the charge cloud to spread, such as thermal diffusion and the repulsion of like charge carriers, could result in some charge carriers spreading into the region between two strips and not being collected. Charge loss can also occur due to crystal impurities which can trap charge carriers as they traverse the detector volume \citep{Knoll,Amman00}.

Charge loss is observed in charge sharing hits, or hits with two adjacent strip hits, on the AC side of the COSI GeDs. Charge loss is not observed on the DC side because the the two sides of the GeDs were processed differently during fabrication. Since the effect is only present on the AC side, we do not need to correct for it during the event calibration; we instead choose the hit energy from the DC side during the strip pairing calibration. Nevertheless, it is important to simulate charge loss because its presence causes more events to be flagged during the strip pairing process. These flagged events are discarded after the event calibration stage of the pipeline.

\begin{figure}
\centering
\subfloat[\label{fig:sdPlotA}]{\includegraphics[width=0.5\textwidth]{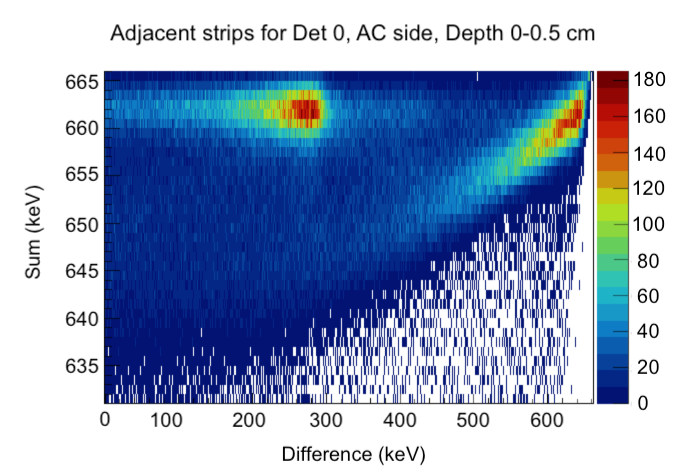}}
\subfloat[\label{fig:sdPlotB}]{\includegraphics[width=0.5\textwidth]{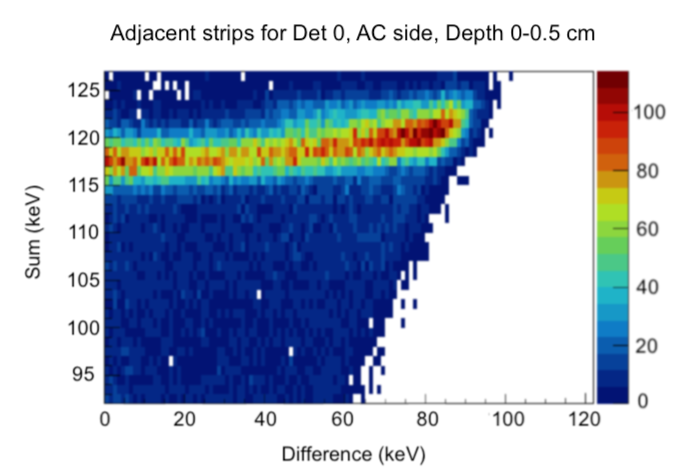}}
\caption{Sum-difference histogram (color online) of two site events from the AC side of detector 0 within a depth of 0-0.5 cm for (a) the 662~keV line of a $^{137}$Cs source and (b) the 122~keV line of a $^{57}$Co source. In (a) charge loss is evident by the hot spot at large differences: if no charge loss were present, the sum of all the counts in this hot spot would be at $\sim$662~keV. In (b), charge loss is evident in the parabolic dip below 122~keV.}
\label{fig:sdPlot}
\end{figure}

To simulate charge loss, we reverse the correction process described in \cite{BandstraThesis} and summarized here. We select hits that contain two adjacent strip hits and relate the sum of the energies $S = E_1+E_2$ of the two adjacent strips to the difference of the energies $D=|E_1-E_2|$. For each hit, ($S$, $D$) is plotted, as in Figure \ref{fig:sdPlot}. At high energies (662 keV, Figure \ref{fig:sdPlotA}), the two hot spot clusters are caused by different physical processes: the cluster at differences of $\sim$250 keV is made up of backscatter events, where the two adjacent strip hits represent two separate interactions (or hits) in the detector with a Compton scatter angle $\phi\sim180^{\circ}$ between the two interactions, while the cluster at differences between \~500 and 662~keV is made up of charge sharing hits. Charge loss is evident by the downward slope in the cluster at large differences, where the sum is less than the line energy of the source (in this case, 662~keV). If no charge loss were present in the detector, the sum would be ideally equal to the line energy.  At low energies, the photons are much less likely to backscatter at $\phi\sim$180$^{\circ}$ due to the kinematics of Compton scattering, and so all low energy hits with two adjacent strip hits are most likely charge sharing hits. Charge loss for low energy $\gamma$-ray sources is evident by the parabolic dip in the sum below the line energy, as in Figure \ref{fig:sdPlotB}.

To simulate the charge loss effect, we consider a phenomenological model for the sum $S$ as a function of difference $D$:

\begin{equation}
\label{eq:chargeLoss}
S(D) = \begin{cases} E_0-\frac{B}{2E_0}(E_0^2-D^2), & E_0 < 300 \text{keV} \\
E_0-B(E_0-D), & E_0 \geq 300 \text{keV} \end{cases}
\end{equation}

\noindent where $E_0$ is the true energy of the hit. The parameter $B$ is the slope of the wing where it meets the line $S(E_0)=E_0$. The model takes into account the linear shape of the outer hot spot at high energies and the curved shape of the hot spot at low energies. $B$ depends slightly on the energy and depth in the detector. We find $B$ as a function of energy for three depth bins in the detector: 0-0.5~cm, 0.5-1~cm, and 1-1.5~cm (the depth is 0~cm at the AC side of the detector and 1.5~cm at the DC side). For each depth bin, we determine $B$ at four different energies (122~keV, 356~keV, 662~keV, and 1333~keV) by fitting the sum-difference plots of four calibration source lines with this model. We then interpolate $B$ linearly as a function of energy to find $B$ at any $E_0$. Once the interpolation function is defined, modeling the charge loss effect is straightforward: as each simulated charge sharing hit initially has a sum $S=E_0$, we apply Equation \ref{eq:chargeLoss} to estimate the $S$ due to charge loss. To assign the hit a reduced energy of $S$, the energy of each contributing strip hit must be reduced. We subtract $(E_0-S)/2$ from the energy of each strip hit to preserve $D$ between the strip hit energies.

We note that although charge sharing can occur across three or more strips, this method of simulating charge loss can only be applied to charge sharing hits containing exactly two adjacent strip hits. An in-depth study of charge loss in charge sharing hits containing more than two adjacent strips similarly to that in \cite{BandstraThesis} has not yet been performed for the COSI detectors. As the majority ($\sim$75\%) of charge sharing hits only contain two adjacent strip hits, however, applying charge loss to charge sharing hits with only two adjacent strips is a reasonable approximation of charge loss in the COSI detectors.

\begin{figure}
\centering
\subfloat[]{\includegraphics[width=0.5\textwidth]{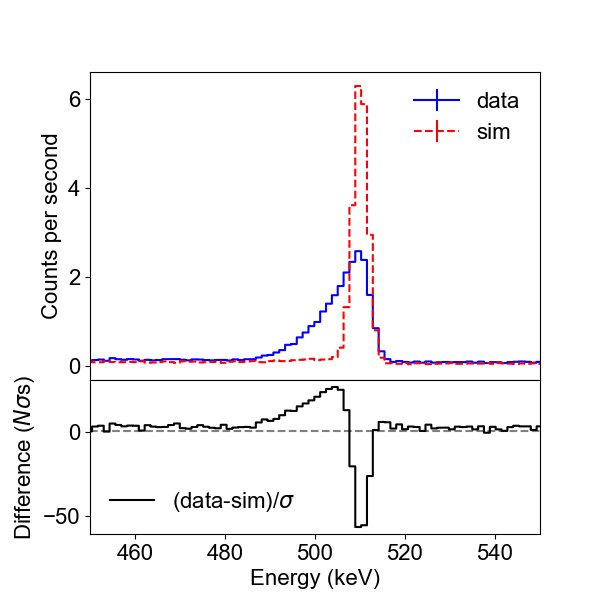}}
\subfloat[]{\includegraphics[width=0.5\textwidth]{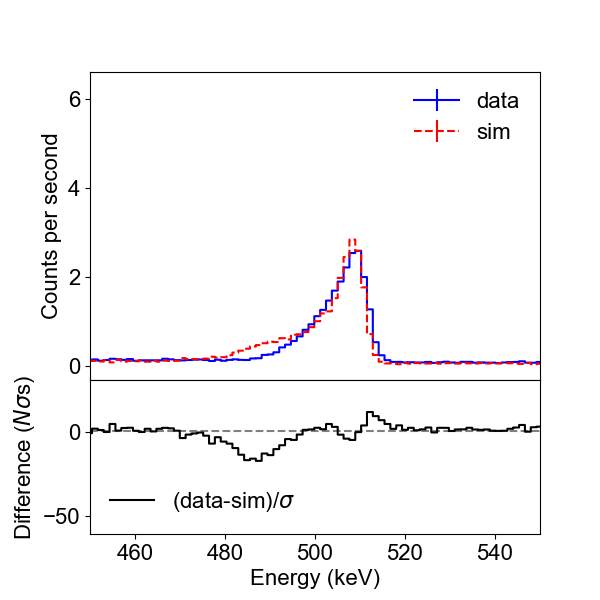}}
\caption{The measured and simulated spectra of the 511 keV line from a $^{22}$Na source (a) before and (b) after charge loss is applied to the simulation. The spectra only contain hits with two adjacent strip hits from the AC side of the detector and one strip hit on the DC side to clearly view the effects of charge loss. Adding charge loss to the simulation significantly improves the agreement between measurements and simulations.}
\label{fig:chargeLoss}
\end{figure}

Figure \ref{fig:chargeLoss} compares the measured and simulated spectra of the 511~keV line from a $^{22}$Na source before and after charge loss is applied to the simulation. These spectra only contain hits with two adjacent strip hits from the AC side of the detector to best see the charge loss effect. After charge loss is included in the simulations, the measured and simulated spectra are in better agreement. The excess tailing on the low-energy side in the simulated spectrum is likely due to the presence of simulated hits with a large $D$. As the strip hit energies for charge sharing hits were determined empirically as described in Section \ref{sec:chargesharing}, a more physical charge sharing simulation could potentially alleviate this excess tailing.

\subsection{Crosstalk}
Crosstalk is the influence of one electronics channel on another. In the COSI GeDs, crosstalk affects hits that contain two adjacent strip hits by amplifying the energy of each strip hit. To simulate crosstalk, we reverse the correction method described in Section \ref{sec:calib} and detailed in \cite{BandstraThesis} and \cite{KieransThesis}. The crosstalk effect is linear, meaning that the energy increase due to crosstalk is proportional to the energy of the original strip plus an offset due to sub-threshold adjacent strips. In other words, if $E_{1,\text{T}}$ and $E_{2,\text{T}}$ are the true energies of the strip hits, the measured energies $E_{1,\text{M}}$ and $E_{2,\text{M}}$ are

\begin{equation}
\label{eq:crosstalk}
\begin{aligned}
E_{1,\text{M}} & = E_{1,\text{T}}+\beta E_{2,\text{T}}-\frac{\alpha}{2} \\
E_{2,\text{M}} & = E_{2,\text{T}}+\beta E_{1,\text{T}}-\frac{\alpha}{2}
\end{aligned}
\end{equation}

\noindent where $\alpha$ and $\beta$ are the sub-threshold offset and correction factor, respectively. These factors are determined for each side of each detector as follows: for multiple calibration sources with known $\gamma$-ray line energies, we construct a spectrum including only hits with two adjacent strip hits on one side of the detector and two non-adjacent strip hits on the other side of the detector. This configuration of strip hits most likely corresponds to two separate interactions in the detector, rather than a charge sharing hit that is adversely affected by charge loss (note that charge sharing hits are affected by crosstalk, but because they are also affected by charge loss, it is best to leave them out when determining $\alpha$ and $\beta$). We fit the peak in each measured spectrum to determine the deviation of the best fit mean energy from the known line energy. We then perform a linear fit of this deviation as a function of energy. The slope of the best fit line is $\beta$ and the offset is $\alpha$. See \cite{BandstraThesis} for more details on determining these factors.

\begin{figure}
\centering
\subfloat[]{\includegraphics[width=0.5\textwidth]{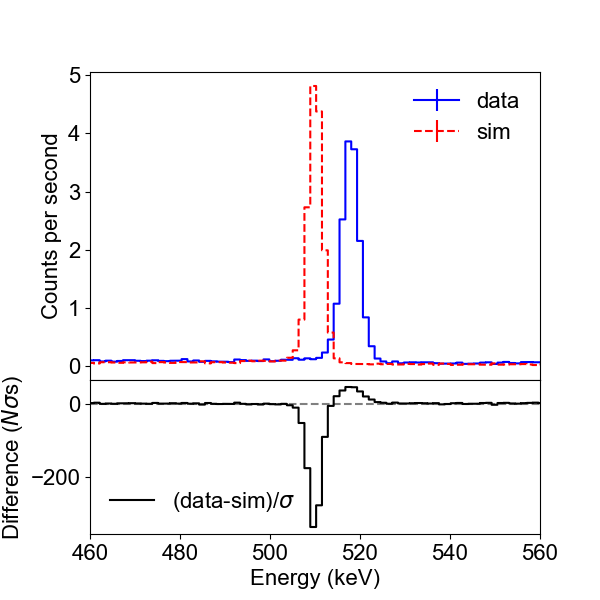}}
\subfloat[]{\includegraphics[width=0.5\textwidth]{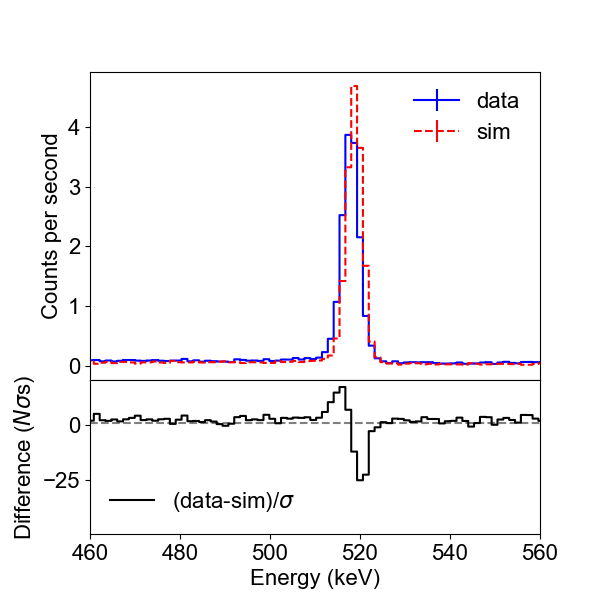}}
\caption{Measured and simulated spectra (a) before and (b) after crosstalk is included in the simulation. The spectra, made before the crosstalk correction, only contain hits with two adjacent strip hits  on the DC side of the detector to clearly view the effect of crosstalk. When crosstalk is included in the simulation, the same line shift is present as in the measurement. Note that crosstalk is corrected in the event calibration pipeline, at which point the line will return to 511~keV.}
\label{fig:crosstalk}
\end{figure}

Once the offset and correction factor are determined, crosstalk can be added to the simulations using Equation \ref{eq:crosstalk}, which estimates the new energy of each strip hit due to crosstalk $E_{n,\text{M}}$ as a function of the original strip hit energy $E_{n,\text{T}}$. Figure \ref{fig:crosstalk} compares the measured and simulated spectra of the 511~keV line from $^{22}$Na before and after crosstalk is applied to the simulation. The spectra only include hits with two adjacent strip hits from the DC side of the detector and two non-adjacent strips on the AC side of the detector to best see the crosstalk effect without interference from the charge loss effect. These spectra are made after the energy calibration but before the strip pairing and crosstalk correction, so the crosstalk effect is still present, causing the line to shift from 511~keV to $\sim$519~keV. The same line shift is present in the simulations after incorporating the crosstalk effect.

\subsection{Energy to pulse height}
We invert the energy calibration (see Section \ref{sec:calib}) to convert energy into pulse height, or ADC value. Each strip has an individual calibration in which the energy is related to the ADC value with a third or fourth order polynomial. We then apply Gaussian noise to the ADC value. For each individual strip, the width of the Gaussian as a function of energy is determined during the energy calibration of that strip.

\subsection{Thresholds and dead strips}
Each strip has two pulse shaping circuits. The ``fast" shaper measures the time of the interaction to an accuracy of 5~ns resolution and has a threshold of about $40-50$~keV. The ``slow" shaper precisely measures the pulse height of the signal (which corresponds to the deposited energy) with minimal noise, and has a threshold of about $20$~keV. Strip hits with energies below the slow threshold are not recorded by the card cage, and strip hits with energies below the fast threshold do not have timing (and therefore depth) information.

\begin{figure}
\centering
\subfloat[\label{fig:lldFstSpecA}]{\includegraphics[width=0.5\textwidth]{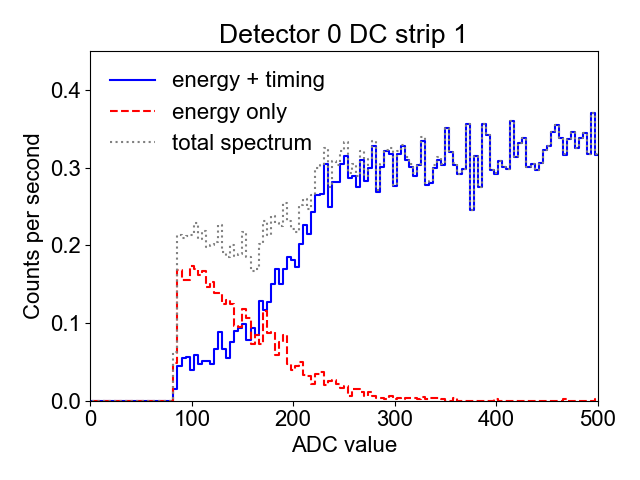}}
\subfloat[\label{fig:lldFstSpecB}]{\includegraphics[width=0.5\textwidth]{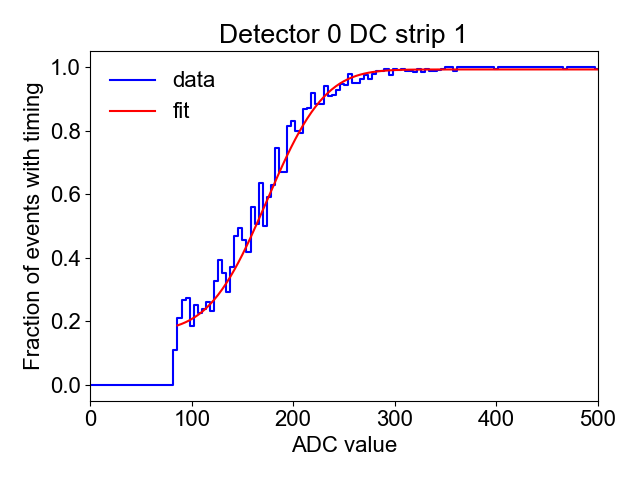}}
\caption{(a) The energy-only and energy-and-timing spectra for a single example strip. The slow threshold results in the sharp cutoff to the spectrum at around ADC value 90. The fast threshold is noisy, so there is no clear cutoff between energy only events and events with energy and timing. (b) The energy-and-timing spectrum divided by the total spectrum for the same strip, representing the fraction of events with timing as a function of ADC value.}
\label{fig:lldFstSpec}
\end{figure}

The threshold values in keV for each strip differ slightly due to noise and gain variations and must be calibrated separately. The slow threshold for each strip is determined by the sharp cutoff in the low-energy spectrum, as shown in Figure \ref{fig:lldFstSpecA}. To apply the slow threshold to the simulations, any strip hits with energy below the slow threshold of that strip are removed.

The energy determination of the fast shaping channel is less precise, so determining the threshold value is more difficult. To do so, we consider two separate spectra for each strip: one containing energy-only events, and one containing events with energy and timing, as shown in Figure \ref{fig:lldFstSpecA}. The spectrum of energy-only events arises because events below the fast threshold lack a timing measurement. The noisiness of the fast shaping channel is evident in Figure \ref{fig:lldFstSpecA} as there is no one point where the energy-only spectrum ends and the energy-and-timing spectrum begins. This noise is a combination of voltage noise, which is Gaussian-distributed, and shot noise from the current, and so the total noise resembles filtered Gaussian noise. Consider dividing the energy-and-timing spectrum by the total spectrum (energy-and-timing added to energy-only), shown in Figure \ref{fig:lldFstSpecB}. For each ADC value, the corresponding y axis value represents the fraction of events that have timing information. These curves can be well fit with an error function, likely due to the Gaussian component of the noise on the fast shaper. This error function is then used to determine whether or not a simulated strip hit should have timing: if a random number drawn between 0 and 1 is greater than the value of the error function at the ADC value of the strip hit, the strip hit's timing is removed.

\begin{figure}
\centering
\subfloat[]{\includegraphics[width=0.5\textwidth]{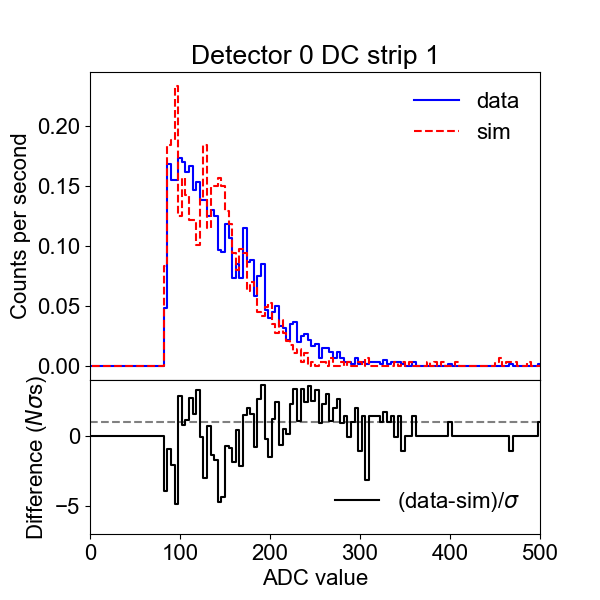}}
\subfloat[]{\includegraphics[width=0.5\textwidth]{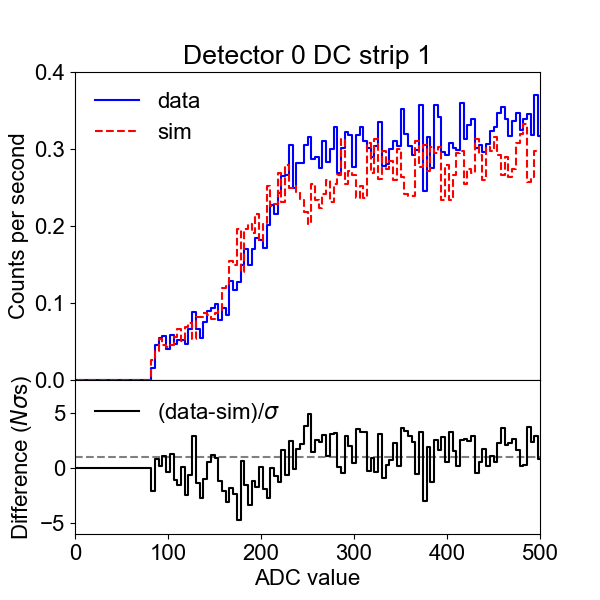}}
\caption{A comparison of the measured and simulated (a) energy-only spectrum and (b) energy-and-timing spectrum for an example strip. There is good agreement between measurements and simulations.}
\label{fig:thresholdResults}
\end{figure}

Figure \ref{fig:thresholdResults} shows a comparison between measured data and simulations for the energy-only spectrum and the energy-and-timing spectrum at low energies. There is fairly good agreement between data and simulations, confirming that our implementation of the thresholds in the DEE is accurate.

A very small fraction of the strips (7 out of 888) are dead. These dead strips could be due to a problem with the detector, the readout electronics, or the signal cable between the two. The DEE discards any simulated strip hits that occur on a dead strip.

\subsection{Guard ring veto}
Each side of each detector has a guard ring that surrounds the strips to prevent surface leakage current from flowing between the anode and cathode strips. The guard ring is also used to veto events in which interactions occur too close to the edge of the detector, where non-uniformities in the electric field can degrade the detector response. If a strip hit occurs on the guard ring in a detector above the guard ring threshold, all other coincident strip hits that occurred in that detector are discarded. To determine the value of the guard ring threshold in keV, we performed an energy calibration of each guard ring channel; the threshold values range from 14~keV to 47~keV, with an average value of 36.6~keV.

\subsection{Trigger conditions}
The card cages only record events that have at least one strip hit above the timing threshold on both sides of the detector. After all effects (except for dead time) have been applied, the DEE ensures that in each triggered detector there is at least one strip hit per side above the timing threshold. If the trigger conditions are not met for a single detector, all strip hits from that detector are removed. Strip hits from other detectors that do meet the trigger conditions are not removed and continue to the next stage.

\subsection{Dead time}The dead time is the time after an event is recorded during which the system cannot record another event. Each card cage is dead for $\sim$10~$\mu$s after a single event occurs in the detector: this is the amount of time it takes for the analog boards to trigger and the coincidence logic to be performed. In the DEE code, if a hit occurs in a certain detector less than 10~$\mu$s after the previous hit, the hit is discarded.

Additional dead time is introduced by the software in the card cages that takes the information about the triggered event from the shapers and parses it into a dataframe packet to send to the flight computer. The card cage takes 625~$\mu$s to process one event, which is determined by inverting the fastest measureable data rate per card cage, measured as 1.6 kHz. Each card cage can process up to 16 events simultaneously. The cause of this dead time is modeled in the DEE as follows: each time a new event occurs in a particular detector and passes the trigger conditions, a timer begins in one available buffer slot. To update the timer, for every new event, the timer is incremented by the time difference between the current and previous event. Once the timer hits 625~$\mu$s, the buffer slot is available again. If an event occurs when there are no available buffer slots, that event is discarded.

\section{Simulation benchmarking}
\label{sec:simulationBenchmarking}
To benchmark the simulations, we compare them to calibration data taken in the lab after the event calibration and event reconstruction steps of the analysis pipeline. We describe these comparisons in this section.

\subsection{Spectrum comparison}

\begin{figure}
\centering
\subfloat[]{\includegraphics[width=0.5\textwidth]{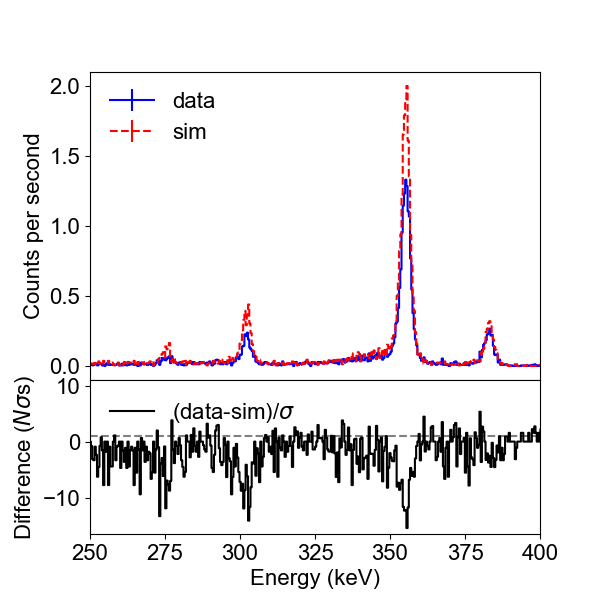}}
\subfloat[]{\includegraphics[width=0.5\textwidth]{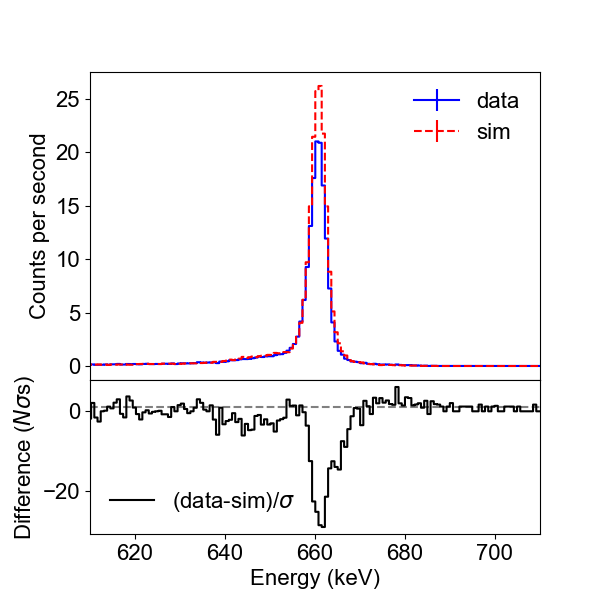}}
\caption{A comparison of the measured and simulated spectrum of (a) a $^{133}$Ba source and (b) a $^{137}$Cs source. The line shapes match very well, but the number of counts in the lines differs.}
\label{fig:specBeforeFF}
\end{figure}

\begin{figure}
\centering
\includegraphics[width=0.7\textwidth]{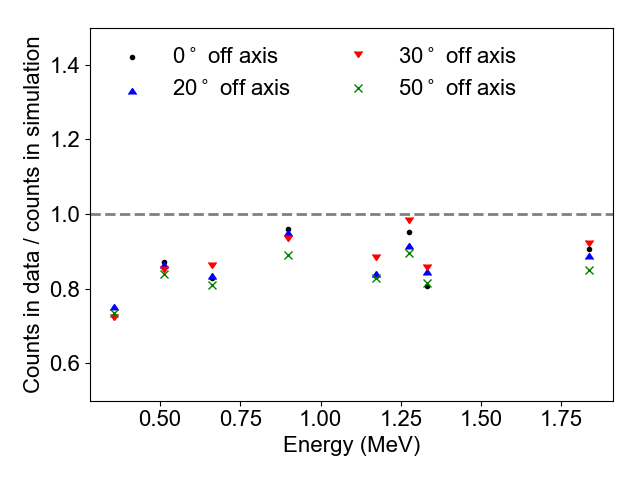}
\caption{The ratio of measured to simulated counts in the lines as a function of energy and off-axis angle.}
\label{fig:peakHeightDiff}
\end{figure}

Figure \ref{fig:specBeforeFF} shows a comparison of the measured and simulated spectra of a $^{133}$Ba and a $^{137}$Cs calibration source. The spectra were made after the event reconstruction step of the analysis pipeline, and therefore include the removal of flagged events from the event calibration. The simulated spectra include all steps of the DEE described in Section \ref{sec:dee}. The measured and simulated spectral continuum and line shapes match very well, illustrating that the DEE is accurately simulating the detector performance. However, there is a discrepancy in the peak height, but not in the continuum, indicating that too many simulated fully absorbed events pass through the pipeline. This discrepancy exists for all calibration sources, as shown in Figure \ref{fig:peakHeightDiff}, which plots the ratio of counts in the measured line to counts in the simulated line as a function of energy and off-axis angle for many calibration sources. The off-axis angle has a weak effect on the differences between measurements and simulations. It is possible that fully absorbed events are affected differently than incompletely absorbed events because of an inaccurate simulation of the guard ring size. Any events containing hits that are vetoed by the guard ring appear to be incompletely absorbed events. A calibration of the guard ring size was not performed before the 2016 flight. If we are assuming a smaller guard ring veto area in the simulations than the true veto area, then the ratio between fully absorbed events and the continuum changes. To correct for this discrepancy, a certain fraction of fully absorbed events $-$ one minus the ratio shown in Figure \ref{fig:peakHeightDiff} $-$ are removed in the DEE. The ratio is linearly interpolated across the energy range. For each fully absorbed event in the DEE, a random number between zero and one is drawn; if that number is greater than the ratio of measured to simulated line counts at the energy of the event, the event is discarded.

\begin{figure}
\centering
\subfloat[]{\includegraphics[width=0.5\textwidth]{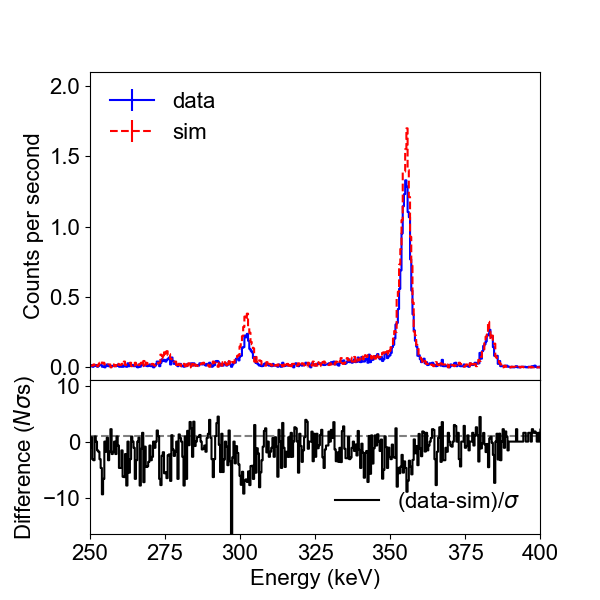}}
\subfloat[]{\includegraphics[width=0.5\textwidth]{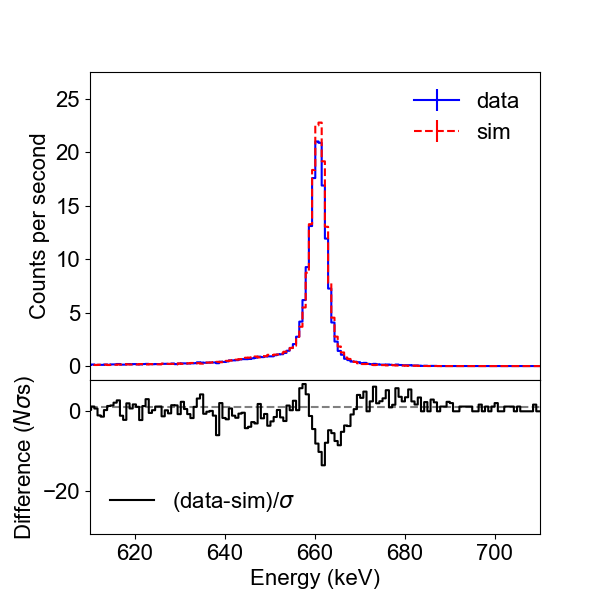}}
\caption{A comparison of the measured and simulated spectra of (a) a $^{133}$Ba calibration source and (b) a $^{137}$Cs calibration source, after some of the fully absorbed events have been removed. The line shapes and heights match very well.}
\label{fig:spectralCompFudgeFactor}
\end{figure}

\begin{figure}
\centering
\includegraphics[width=0.7\textwidth]{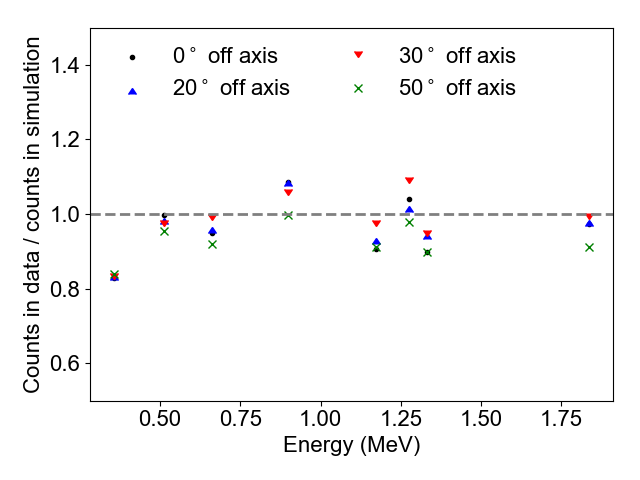}
\caption{A comparison of the ratio of measured to simulated counts in the lines as a function of energy and off-axis angle, after some of the fully absorbed events are removed from the simulations.}
\label{fig:peakHeightDiffFudgeFactor}
\end{figure}

Figure \ref{fig:spectralCompFudgeFactor} shows a comparison of the measured and simulated spectrum of two calibration sources after removing a fraction of the fully absorbed events, as described above, and the peak heights match very well. Figure \ref{fig:peakHeightDiffFudgeFactor} shows an updated plot of the ratio of counts in the measured line to counts in the simulated line as a function of energy and off-axis angle; the discrepancy of peak heights is now minimized.

\begin{figure}
\centering
\includegraphics[width=0.7\textwidth]{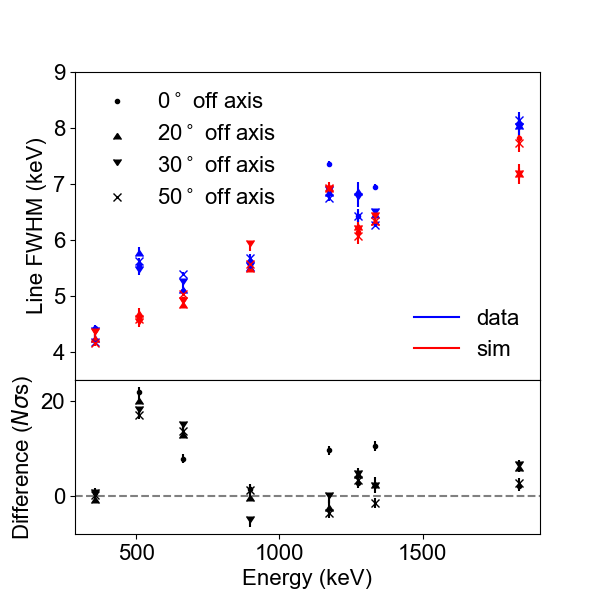}
\caption{A comparison of the measured and simulated full width at half maximum (FWHM) of each calibration line energy as a function of energy and off-axis angle. The relatively large difference between the measured and simulated FWHM at 511~keV is due to the assumption in the simulations that the electron-positron pair that produces the 511~keV $\gamma$-rays is at rest, when in reality the electron and positron have some momentum; this non-zero momentum leads to the broadening of the line.}
\label{fig:specFWHM}
\end{figure}

To compare the shape of the measured and simulated spectra, we fit each calibration source energy line with a Gaussian function. The results are shown in Table \ref{tab:centroidFWHM}, and the full width at half maximum (FWHM) of each line as a function of energy and off-axis angle is plotted in Figure \ref{fig:specFWHM}. The relatively large difference between the measured and simulated FWHM at 511~keV is due to the fact that the simulations assume that the electron-positron pair that annihilates to produce the 511~keV $\gamma$-rays is at rest, when in reality the electron and positron have some momentum, which results in the broadening of the line. The smaller deviations between the measured and simulated FWHM could be the result of imperfections in the energy calibration and crosstalk correction present in the measured data. Note that we can assume that the energy calibration and crosstalk correction are perfect for the simulations because the reversal of these calibrations is applied to the simulated events as part of the DEE.

\begin{table}
\centering
\footnotesize
\begin{tabular}{c|c|cc|cc}
Energy & Off-axis angle ($^\circ$) & \multicolumn{2}{c|}{Measurement} & \multicolumn{2}{c}{Simulation} \\
\emph{Source} & & Centroid (keV) & FWHM (keV) & Centroid (keV) & FWHM (keV) \\ \hline\hline
356.01 & 0 & $355.36\pm0.02$ & $4.43\pm0.05$ & $355.35\pm0.02$ & $4.39\pm0.05$ \\
\emph{$^{133}$Ba} & 20 & $355.37\pm0.02$ & $4.28\pm0.05$ & $355.36\pm0.02$ & $4.29\pm0.05$ \\
 & 30 & $355.34\pm0.03$ & $4.34\pm0.06$ & $355.35\pm0.02$ & $4.33\pm0.05$ \\
 & 50 & $355.37\pm0.02$ & $4.17\pm0.06$ & $355.38\pm0.02$ & $4.16\pm0.05$ \\ \hline
511.0 & 0 & $510.3\pm0.02$ & $5.56\pm0.04$ & $510.38\pm0.02$ & $4.61\pm0.04$ \\
\emph{$^{22}$Na} & 20 & $510.29\pm0.02$ & $5.82\pm0.05$ & $510.38\pm0.02$ & $4.74\pm0.04$ \\
 & 30 & $510.21\pm0.02$ & $5.44\pm0.05$ & $510.4\pm0.02$ & $4.5\pm0.05$ \\
 & 50 & $510.15\pm0.03$ & $5.62\pm0.06$ & $510.38\pm0.02$ & $4.59\pm0.05$ \\ \hline
661.66 & 0 & $660.87\pm0.01$ & $5.1\pm0.02$ & $661.02\pm0.01$ & $4.95\pm0.02$ \\
\emph{$^{137}$Cs} & 20 & $660.66\pm0.01$ & $5.17\pm0.02$ & $661.02\pm0.01$ & $4.89\pm0.02$ \\
 & 30 & $660.69\pm0.01$ & $5.21\pm0.02$ & $661.04\pm0.01$ & $4.88\pm0.02$ \\
 & 50 & $660.78\pm0.01$ & $5.4\pm0.03$ & $661.03\pm0.01$ & $5.04\pm0.02$ \\ \hline
898.04 & 0 & $897.17\pm0.02$ & $5.63\pm0.07$ & $897.37\pm0.03$ & $5.53\pm0.07$ \\
\emph{$^{88}$Y} & 20 & $897.07\pm0.03$ & $5.55\pm0.07$ & $897.44\pm0.03$ & $5.54\pm0.07$ \\
 & 30 & $897.08\pm0.03$ & $5.5\pm0.07$ & $897.34\pm0.03$ & $5.88\pm0.08$ \\
 & 50 & $896.97\pm0.03$ & $5.67\pm0.08$ & $897.44\pm0.03$ & $5.58\pm0.08$ \\ \hline
1173.23 & 0 & $1172.15\pm0.02$ & $7.36\pm0.05$ & $1172.57\pm0.01$ & $6.85\pm0.05$ \\
\emph{$^{60}$Co} & 20 & $1172.08\pm0.02$ & $6.89\pm0.05$ & $1172.56\pm0.02$ & $6.99\pm0.05$ \\
 & 30 & $1172.16\pm0.02$ & $6.86\pm0.05$ & $1172.58\pm0.02$ & $6.88\pm0.06$ \\
 & 50 & $1172.24\pm0.02$ & $6.74\pm0.05$ & $1172.63\pm0.02$ & $6.94\pm0.06$ \\ \hline
1274.54 & 0 & $1273.62\pm0.04$ & $6.42\pm0.1$ & $1273.83\pm0.04$ & $6.13\pm0.11$ \\
\emph{$^{22}$Na} & 20 & $1273.26\pm0.05$ & $6.92\pm0.13$ & $1274.0\pm0.04$ & $6.28\pm0.12$ \\
 & 30 & $1273.31\pm0.05$ & $6.73\pm0.13$ & $1273.9\pm0.05$ & $6.16\pm0.13$ \\
 & 50 & $1273.22\pm0.05$ & $6.44\pm0.11$ & $1273.98\pm0.04$ & $6.06\pm0.13$ \\ \hline
1332.49 & 0 & $1331.43\pm0.02$ & $6.95\pm0.05$ & $1331.89\pm0.02$ & $6.45\pm0.04$ \\
\emph{$^{60}$Co} & 20 & $1331.37\pm0.02$ & $6.5\pm0.04$ & $1331.85\pm0.02$ & $6.37\pm0.05$ \\
 & 30 & $1331.43\pm0.02$ & $6.46\pm0.05$ & $1331.88\pm0.02$ & $6.39\pm0.05$ \\
 & 50 & $1331.5\pm0.02$ & $6.27\pm0.05$ & $1331.95\pm0.02$ & $6.34\pm0.05$ \\ \hline
1836.05 & 0 & $1834.78\pm0.05$ & $8.08\pm0.12$ & $1835.61\pm0.05$ & $7.83\pm0.15$ \\
\emph{$^{88}$Y} & 20 & $1834.42\pm0.05$ & $8.1\pm0.13$ & $1835.49\pm0.04$ & $7.23\pm0.13$ \\
 & 30 & $1834.4\pm0.05$ & $7.98\pm0.14$ & $1835.57\pm0.05$ & $7.15\pm0.14$ \\
 & 50 & $1834.48\pm0.06$ & $8.14\pm0.15$ & $1835.64\pm0.05$ & $7.73\pm0.15$ \\ \hline
 \end{tabular}
\caption{A comparison of the line widths and centroids of a Gaussian fit of each calibration source line for measurements and simulations.}
\label{tab:centroidFWHM}
\end{table}

\begin{figure}
\centering
\subfloat[]{\includegraphics[width=0.5\textwidth]{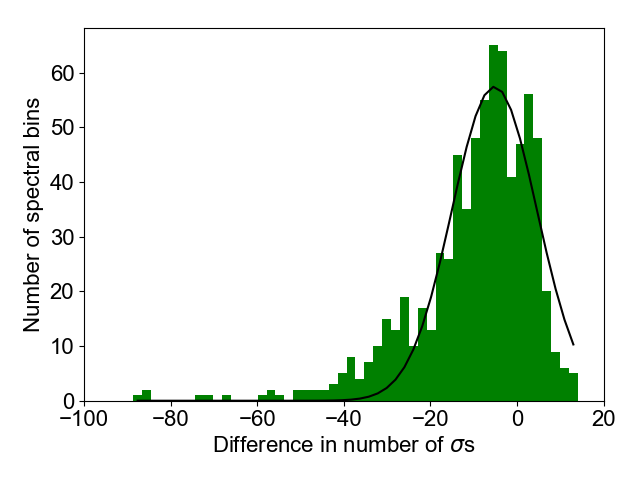}}
\subfloat[]{\includegraphics[width=0.5\textwidth]{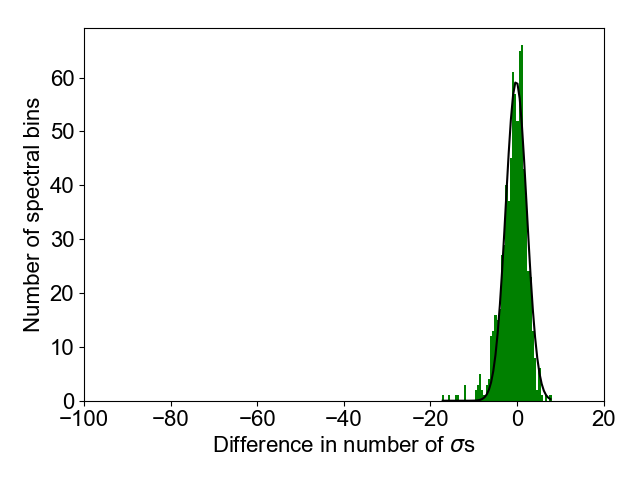}}
\caption{The distribution of $z=(N_\text{M}-N_\text{S})/\sigma_\text{M}$ when (a) the simulations are processed with the standard MEGAlib DEE and (b) the simulations are processed with the COSI-specific DEE. The match between measurements and simulations is much better when the simulations are processed with the COSI-specific DEE: the distribution is about four times narrower and centered near zero.}
\label{fig:statSpecComp}
\end{figure}

To quantitatively assess the improvements we have made with the COSI-specific DEE, we compare the measurements to the simulations as follows: for each bin in the spectrum, we compute the difference in number of sigmas $z=(N_\text{M} - N_\text{S})/\sigma_\text{M}$, where $N_\text{M}$ and $N_\text{S}$ are number of measured and simulated counts in the bin and $\sigma_\text{M}$ is the error on $N_\text{M}$ (with photon counting statistics, $\sigma_\text{M} = \sqrt{N_\text{M}}$). We then make a histogram of the $z$ values for each bin in the spectrum and fit the resulting distribution with a Gaussian function, as in Figure \ref{fig:statSpecComp}. If the measurements and simulations are in good agreement with each other, the mean of the Gaussian fit will be close to zero and the FWHM of the Gaussian fit will be small. Note that we include the entire spectral continuum in this analysis as well as the lines; it is important for both to match well.

We performed this analysis for two cases: first, the simulations were processed with the standard MEGAlib DEE, and second, the simulations were processed with the COSI-specific DEE described in Section \ref{sec:dee} of this paper. Figure \ref{fig:statSpecComp} compares the distributions of $z$ values for these two cases, using a calibration run of a $^{137}$Cs source directly on-axis. The distribution is four times narrower and the mean of the distribution is five standard deviations closer to zero when we use the COSI-specific DEE compared to when we use the standard MEGAlib DEE. Table \ref{tab:spectraMatch} compares the Gaussian fit parameters for a number of different spectra as a function of calibration source and off-axis angle. On average, the FWHM is 2.1 times smaller and the mean is 1.8 standard deviations closer to zero when we use the COSI-specific DEE, indicating that we have achieved better agreement between measurements and simulations.

\begin{table}
\centering
\footnotesize
\begin{tabular}{c|c|cc|cc}
Source & Off-axis angle ($^\circ$) & \multicolumn{2}{c|}{Standard MEGAlib DEE} & \multicolumn{2}{c}{COSI-specific DEE} \\
\emph{Energies (keV)} & & Mean & FWHM & Mean & FWHM \\ \hline\hline
$^{133}$Ba & 0 & -3.7 & 18.9 & -1.1 & 7.17 \\                                               
\emph{276.04, 302.85} & 20 & -3.1 & 17.3 & -0.38 & 6.82 \\                         
\emph{356.01, 383.83} & 30 & -1.8 & 16.5 & -0.58 & 5.69 \\                         
 & 50 & -3.1 & 11.4 & -0.43 & 5.82 \\ \hline                                                    
$^{22}$Na & 0 & 1.1 & 3.49 & 0.29 & 4.42 \\                                             
\emph{511.00} & 20 & 0.53 & 4.32 & 0.23 & 4.2 \\                          
\emph{1274.54} & 30 & 0.8 & 3.88 & 0.26 & 4.02 \\                                      
 & 50 & 0.52 & 4.33 & 0.33 & 3.83 \\ \hline                                                    
$^{137}$Cs & 0 & -5.2 & 23.2 & -1.8 & 5.97 \\                                              
\emph{661.66} & 20 & -5.3 & 20.3 & -1.8 & 5.2 \\                                         
 & 30 & -4.7 & 16.7 & -1.5 & 5.43 \\                                                                
 & 50 & -5.9 & 15.1 & -2.0 & 5.15 \\ \hline                                                     
$^{88}$Y & 0 & 0.93 & -3.57 & 0.47 & 3.59 \\                                                
\emph{898.04} & 20 & 0.86 & -3.76 & 0.34 & 3.56 \\                                    
\emph{1836.05} & 30 & 0.71 & -3.87 & 0.23 & 3.57 \\                                   
 & 50 & 0.43 & 3.98 & 0.09 & 3.42 \\ \hline
$^{60}$Co & 0 & -1.7 & 8.37 & -0.4 & 5.38 \\
\emph{1173.23} & 20 & -1.0 & 7.24 & -0.36 & 5.2 \\
\emph{1332.49} & 30 & -0.74 & 6.61 & -0.29 & 4.8 \\
 & 50 & -2.2 & 6.96 & -0.72 & 4.99 \\ \hline
\end{tabular}
\caption{A comparison of how well the measured and simulated spectra match for a variety of calibration sources and off-axis angles. We compare the measurements to simulations processed with the standard MEGAlib DEE and to simulations processed with the COSI-specific DEE. We show the mean and FWHM of the Gaussian fit to the distribution of the difference in number of sigmas $z=(N_\text{M}-N_\text{S})/\sigma_\text{M}$, where $z$ is computed for each spectral bin.}
\label{tab:spectraMatch}
\end{table}

When analyzing COSI data of astrophysical sources, we compute the instrument response from simulations. Thus it is important to determine the systematic error that comes from the remaining discrepancies between measurements and simulations that we must apply to quantities such as the measured flux. To do so, for each calibration source line energy, we calculate the integral of the measured and simulated spectra between $E_\text{centroid}-2\sigma$ and $E_\text{centroid}+2\sigma$, where $E_\text{centroid}$ is the center and $\sigma$ is the width of the Gaussian fit to the line (see Table \ref{tab:centroidFWHM} for $E_\text{centroid}$ and $\sigma$ values). We then determine the systematic error required such that the number of measured counts $N_M$ in the line is consistent with the number of simulated counts $N_S$ in the line. The results are shown in Table \ref{tab:sysErrSpec}, where the error bars on the systematic errors are the statistical errors of the number of measured counts, or $\sqrt{N_M}$. We performed this analysis for four off-axis angles in COSI's field of view (0$^\circ$, 20$^\circ$, 30$^\circ$, and 50$^\circ$), and the results shown in Table \ref{tab:sysErrSpec} are the results averaged over the four off-axis angles. The average magnitude of the systematic error is 6.3\%, well within our target accuracy of 10\%.

\begin{table}
\centering
\footnotesize
\begin{tabular}{c|c|c}
Energy (keV) & Source & Systematic error (\%) \\ \hline\hline
356.01 & $^{133}$Ba & $+20.2\pm0.3$ \\
511.00 & $^{22}$Na & $+1.6\pm0.2$ \\
661.66 & $^{137}$Cs & $+4.4\pm0.1$ \\
898.04 & $^{88}$Y & $-5.3\pm0.3$ \\
1173.23 & $^{60}$Co & $+6.3\pm0.1$ \\
1274.54 & $^{22}$Na & $-4.8\pm0.4$ \\
1332.49 & $^{60}$Co & $+7.2\pm0.2$ \\
1836.05 & $^{88}$Y & $-0.3\pm0.4$ \\
\end{tabular}
\caption{The systematic error that must be applied to the measured flux, due to the remaining discrepancies between measurements and simulations.}
\label{tab:sysErrSpec}
\end{table}

\subsection{Angular resolution measure comparison}
\begin{figure}
\centering
\subfloat[]{\includegraphics[width=0.4\textwidth]{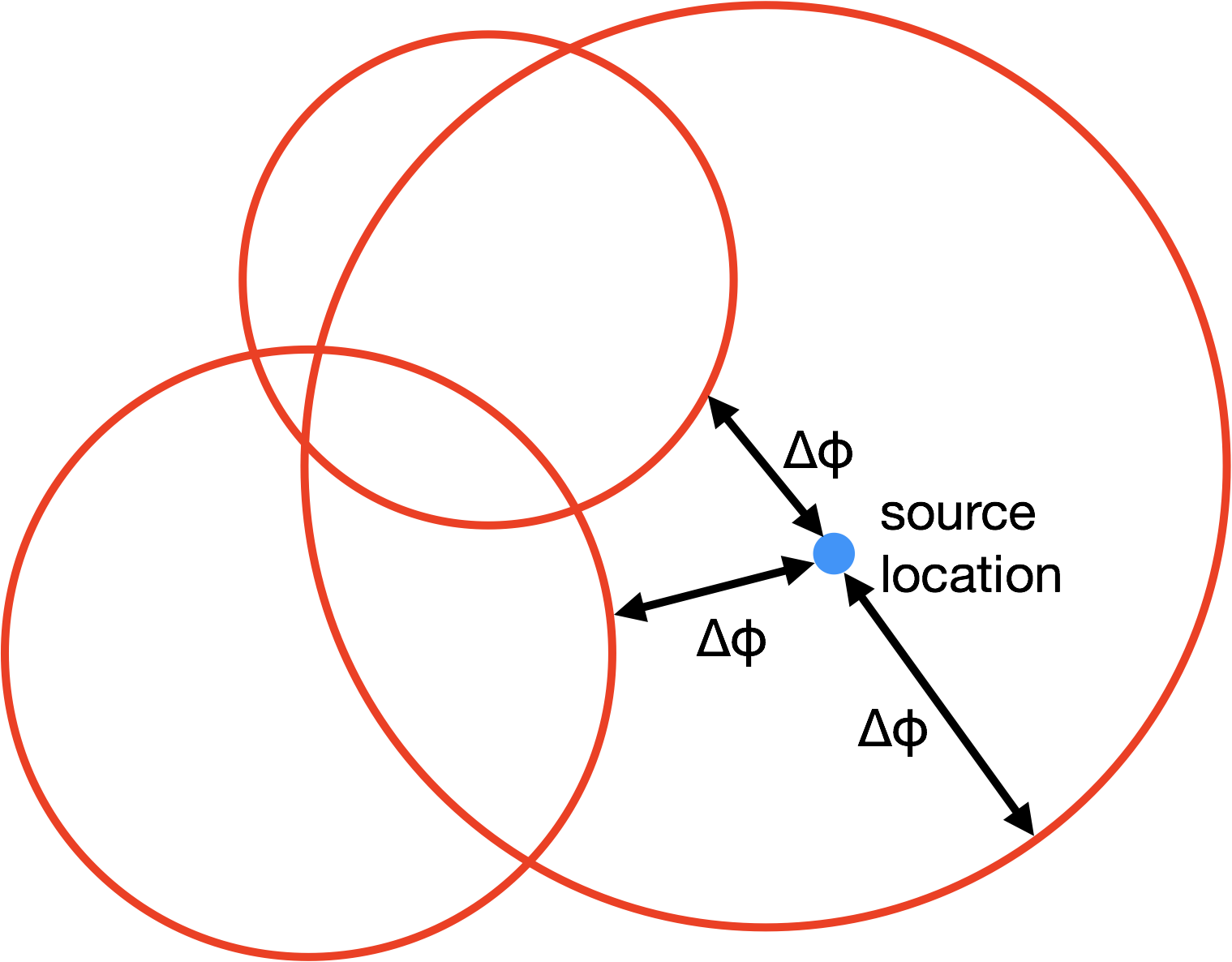}}
\qquad
\subfloat[]{\includegraphics[width=0.5\textwidth]{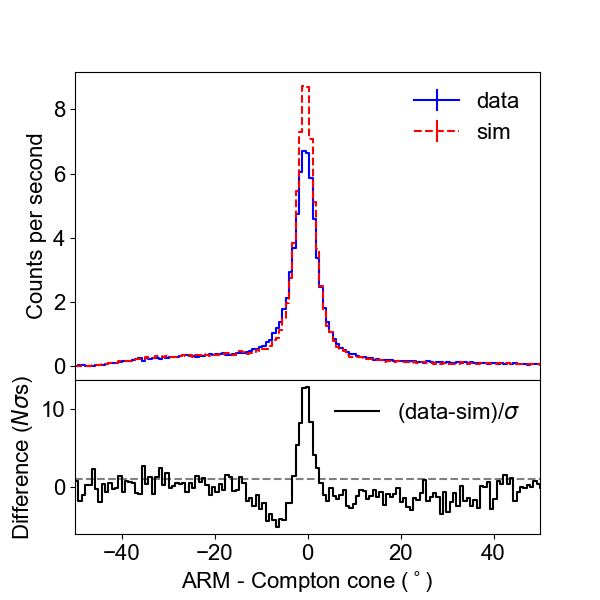}}
\caption{(a) A schematic demonstrating how the ARM is calculated: the blue circle represents the source position and each red circle represents a single photon's Compton circle. The ARM distribution is a histogram of the $\Delta\phi$ values for each photon. (b) A comparison of the measured and simulated ARM distributions of the 662~keV line of a $^{137}$Cs source directly on-axis.}
\label{fig:ARMexplanation}
\end{figure}

The angular resolution measure (ARM) is used to characterize the spatial resolving capabilities of a Compton telescope and is a useful benchmarking tool, as it depends on the energy and position resolution and the results of the event reconstruction. The ARM is the distribution of the smallest angular distance between the known origin of the photons and each Compton circle (see the inset in Figure \ref{fig:ARMexplanation}). If the photon origin falls outside the Compton circle, the angular distance is positive, whereas if it falls inside the Compton circle, the angular distance is negative. The FWHM of the ARM defines the angular resolution of the instrument. Figure \ref{fig:ARMexplanation} shows a comparison of the measured and simulated ARM distributions for the 662 keV line of a $^{137}$Cs source.

Table \ref{tab:armMatch} compares the Gaussian fit parameters of the distributions of $z$ values computed by comparing the measured and simulated ARM distribution. Again, we compare the simulations processed with the standard MEGAlib DEE to the simulations processed with the COSI-specific DEE. Since the shape of the ARM distribution is dependent on energy, we make a separate ARM distribution for each calibration source line energy. On average, the FWHM of the Gaussian fits to the $z$ distribution is 2.3 times smaller and the mean is 1.6 standard deviations closer to zero when the simulations are processed with the COSI-specific DEE, indicating that we have achieved better agreement between measurements and simulations in the ARM distributions as well as in the spectra.

\begin{table}
\centering
\footnotesize
\begin{tabular}{c|c|cc|cc}
Energy (keV) & Off-axis angle ($^\circ$) & \multicolumn{2}{c|}{Standard MEGAlib DEE} & \multicolumn{2}{c}{COSI-specific DEE} \\
\emph{Source} & & Mean & FWHM & Mean & FWHM \\ \hline\hline
356.01 & 0 & 1.9 & 16.1 & 0.26 & 4.46 \\
\emph{$^{133}$Ba} & 20 & 4.1 & 14.4 & 0.17 & 2.93 \\
 & 30 & 0.2 & 5.09 & 0.57 & 4.33 \\
 & 50 & 42.4 & 35.3 & -0.09 & 7.09 \\ \hline
511.0 & 0 & -0.1 & 6.62 & 0.69 & 2.94 \\
\emph{$^{22}$Na} & 20 & -1.3 & 4.52 & 0.65 & 2.77 \\
 & 30 & -0.24 & 4.66 & 0.57 & 3.07 \\
 & 50 & -1.1 & 5.47 & 0.26 & 3.2 \\ \hline
661.66 & 0 & -2.7 & 13.1 & 0.99 & 4.15 \\
\emph{$^{137}$Cs} & 20 & -0.95 & 9.7 & 0.73 & 3.34 \\
 & 30 & -1.0 & 6.16 & 0.67 & 3.56 \\
 & 50 & -1.1 & 6.14 & 0.03 & 3.59 \\ \hline
898.04 & 0 & 0.8 & 6.72 & 0.94 & 3.21 \\
\emph{$^{88}$Y} & 20 & 0.56 & 3.12 & 0.79 & 2.94 \\
 & 30 & 0.59 & 3.53 & 0.57 & 3.01 \\
 & 50 & 0.54 & 3.13 & 0.82 & 2.59 \\ \hline
1173.23 & 0 & -1.2 & 9.22 & -1.2 & 5.6 \\
\emph{$^{60}$Co} & 20 & -0.27 & 6.79 & -0.53 & 3.98 \\
 & 30 & -0.35 & -4.53 & -0.84 & 3.52 \\
 & 50 & -0.41 & 3.54 & -0.83 & 3.63 \\ \hline
1274.54 & 0 & 0.81 & 4.04 & 0.69 & 2.62 \\
\emph{$^{22}$Na} & 20 & 0.66 & 3.23 & 0.65 & 2.51 \\
 & 30 & 0.8 & 4.46 & 0.9 & 3.9 \\
 & 50 & 0.51 & 3.16 & 0.51 & 2.72 \\ \hline
1332.49 & 0 & -0.75 & 6.56 & 0.25 & 4.3 \\
\emph{$^{60}$Co} & 20 & -0.43 & 3.45 & 0.41 & 4.0 \\
 & 30 & -0.58 & 4.65 & 0.23 & 3.3 \\
 & 50 & -0.49 & 3.6 & 0.17 & 3.61 \\ \hline
1836.05 & 0 & 0.31 & 4.9 & 0.7 & 3.08 \\
\emph{$^{88}$Y} & 20 & 0.6 & 3.77 & 0.61 & 3.05 \\
 & 30 & 0.53 & 3.11 & 0.65 & 2.39 \\
 & 50 & 0.34 & 2.6 & 0.28 & 3.06 \\ \hline
\end{tabular}
\caption{A comparison of how well the measured and simulated ARM histograms match for a variety of calibration sources and off-axis angles. We compare the measurements to simulations processed with the standard MEGAlib DEE and to simulations processed with the COSI-specific DEE. We show the mean and FWHM of the Gaussian fit to the distribution of the difference in number of sigmas $z=(N_\text{M}-N_\text{S})/\sigma_\text{M}$, where $z$ is computed for each bin in the ARM histogram.}
\label{tab:armMatch}
\end{table}

\begin{figure}
\centering
\includegraphics[width=0.7\textwidth]{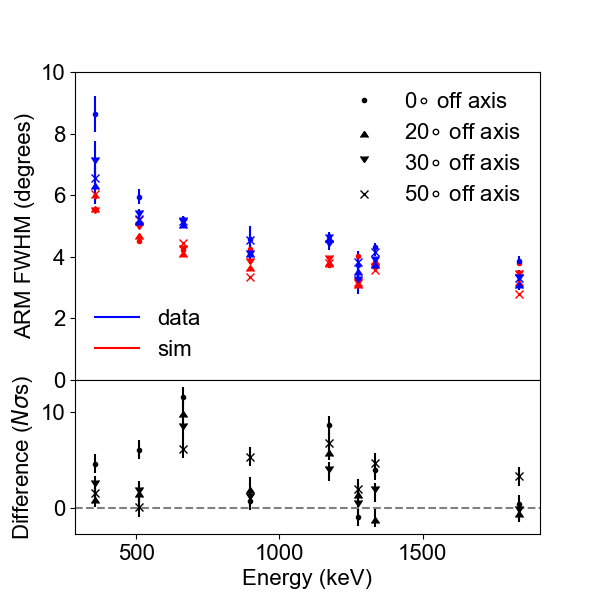}
\caption{A comparison of the measured and simulated ARM FWHM as a function of energy.}
\label{fig:ARMvEnergy}
\end{figure}

\begin{figure}
\centering
\subfloat[]{\includegraphics[width=0.5\textwidth]{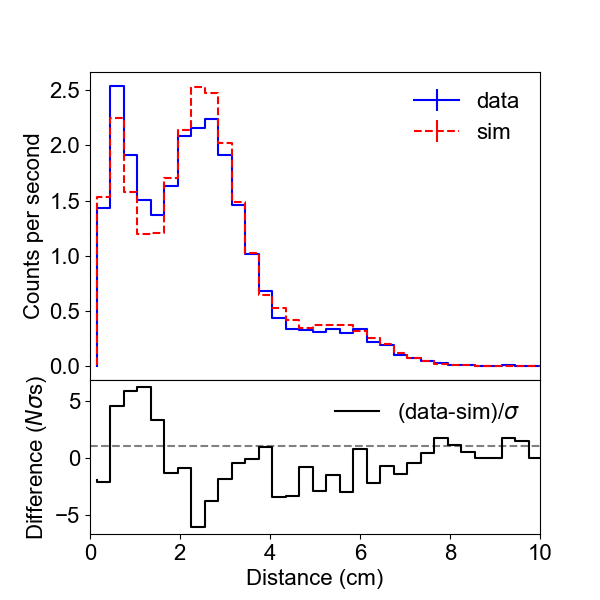}}
\subfloat[]{\includegraphics[width=0.5\textwidth]{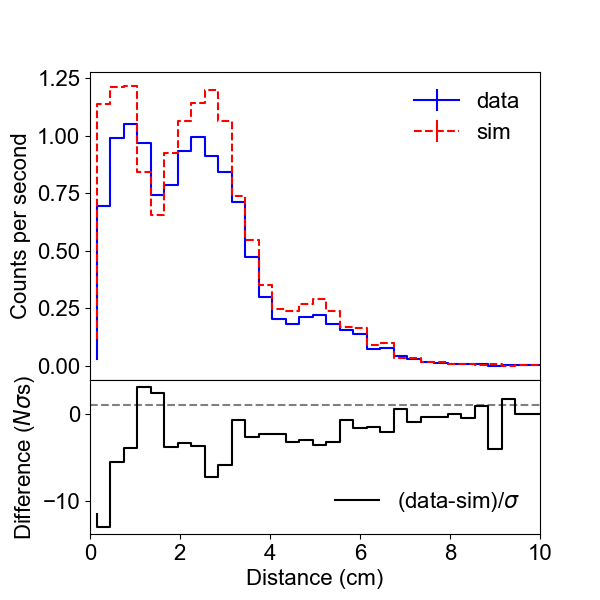}}
\caption{A comparison of the measured and simulated distribution of distances between the first and second interaction for (a) the 511~keV line of a $^{22}$Na source and (b) the 1.836~MeV line of a $^{88}$Y source. The discrepancies below 2.5~cm contribute to the differences between the measured and simulated the ARM FWHM.}
\label{fig:firstiadistance}
\end{figure}

Figure \ref{fig:ARMvEnergy} compares the measured and simulated ARM FWHM as a function of energy and off-axis angle. The measured and simulated ARM FWHM values are fairly close with an average residual of 4.4$\sigma$, but better agreement is desirable. One of the major contributions to the differences  between the measured and simulated ARM FWHM is the discrepancy in the distribution of distances between the first two interactions, and discrepancies in the Compton scatter angle distribution at low energies. Figure \ref{fig:firstiadistance} shows the measured and simulated distribution of distances between the first two interactions at 511~keV and 1836~keV. At distances smaller than $\sim$2.5~cm, the shape of the measured and simulated distributions differ significantly. At low energies, there are more measured events at small distances, and at high energies, the simulated distribution appears offset compared to the measured distribution. Selecting events with a distance between the first two interactions greater than 2.5~cm leads to better agreement between the measured and simulated ARM FWHM values, with an average residual of 1.9$\sigma$.

It is unclear why this discrepancy occurs in the distance distributions. We note that distances smaller than 2.5~cm are likely due to multiple hits in the same detector, whereas larger distances are likely due to multiple hits in different detectors. One possible explanation for the observed discrepancy is that the strip pairing algorithm groups what should be two simulated hits as one. This could cause a dearth of events with a small interaction distance between the first two hits. This issue could potentially be improved by more physical models of charge sharing and charge loss, which requires a detailed charge transport simulation and is thus beyond the scope of this work. Additionally, the differences in the measured and simulated depth distributions of charge sharing hits (Figure \ref{fig:depthDist}) should add discrepancies to the overall distance distributions. Improvements to the depth calibration algorithm could potentially resolve some of the discrepancies in the distance distributions.

\begin{figure}
\centering
\subfloat[]{\includegraphics[width=0.5\textwidth]{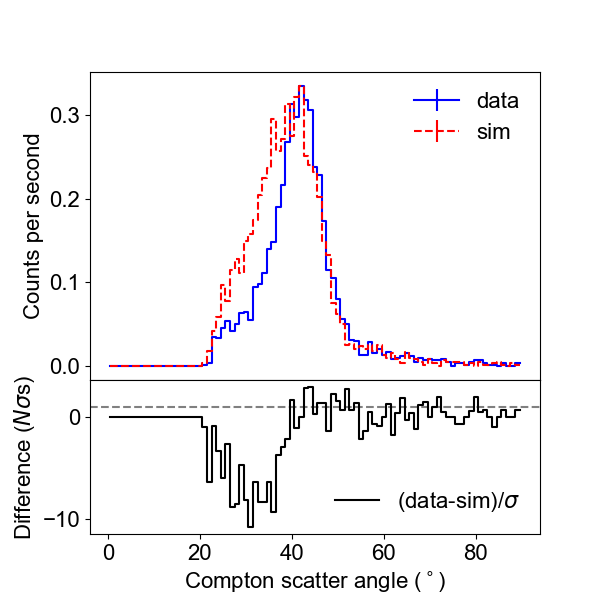}}
\subfloat[]{\includegraphics[width=0.5\textwidth]{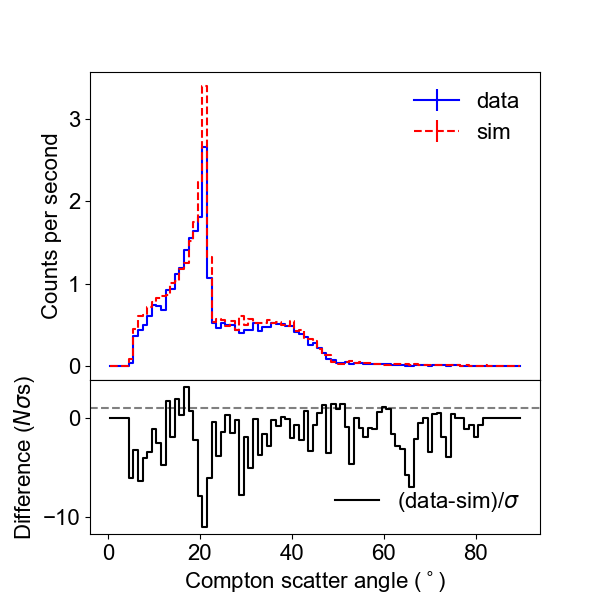}}
\caption{A comparison of the measured and simulated  distribution of initial Compton scatter angles for (a) the 356 keV line of a $^{133}$Ba source and (b) the 1.3~MeV line of a $^{60}$Co source. These distributions agree better at high energies than at low energies. The measured and simulated ARM FWHM are in better agreement at low energies when only events with Compton scatter angles $>40^\circ$ are used.}
\label{fig:scatterangles}
\end{figure}

Figure \ref{fig:scatterangles} compares the measured and simulated initial Compton scatter angle ($\phi$ in Figure \ref{fig:cosiExp}) distribution at 356~keV and 1333~keV. The discrepancies between the measured and simulated scatter angle distribution at low energies is another contributing factor to the differences in the ARM FWHM. Selecting events with Compton scatter angles above 40$^\circ$ and distances greater than 2.5~cm results in excellent agreement between the measured and simulated ARM FWHM values at 356~keV, with a residual of 0.95$\sigma$ (the residual at 356~keV is 4.6$\sigma$ with open selections and 2.6$\sigma$ when selecting events with distances greater than 2.5~cm). The observed difference in Compton scatter angle distribution at low energies could potentially be related to the distance and depth discrepancies. Low energy events are more likely to only interact twice in the detector, and these two-site events are difficult to properly reconstruct. Any systematic error on the interaction distance is likely to translate to an incorrect Compton scatter angle when reconstructing two-site events, which could explain the observed discrepancies in Compton scatter angle distributions. However, it is unclear why smaller Compton scatter angles are affected more than larger Compton scatter angles.

\section{Conclusions}
The COSI-specific DEE described here is an important addition to the COSI analysis pipeline. The instrument response for imaging, spectral, and polarization analysis is computed from simulations, and so it is imperative that the simulations closely resemble the measured COSI data. Careful modeling of the effects in the COSI detectors and readout electronics brings us significantly closer to this goal. We have compared the simulations to calibrations at various stages throughout the analysis pipeline to ensure that they match well. Our comparisons of the spectra and ARM after the event reconstruction indicate very good agreement between measurements and simulations in comparison to the standard MEGAlib DEE. We have determined the systematic error that must be applied to measured fluxes due to remaining discrepancies between measurements and simulations. The systematic error for each line is less than 15\%, and the average of the magnitude of the systematic error is 6.3\%. Better agreement could potentially be achieved with more physical models of the detector effects, which requires a detailed charge transport simulation. Work on a detailed charge transport simulation for the COSI GeDs is currently ongoing.

The instrument response computed from these simulations is a key step in allowing us to accurately analyze data from the 2016 COSI flight and any future COSI flights, thus facilitating advances in understanding astrophysical sources of soft $\gamma$-rays.

\section{Acknowledgements}
We thank Brent Mochizuki for his detailed explanations of the card cage FPGA algorithms and Steve McBride for his help in understanding the analog electronics in the card cages. Support for COSI is provided by NASA Astrophysics Research and Analysis grant NNX14AC81G. C. S. is supported by a NASA Earth and Space Sciences Fellowship. T. S. is supported by the German Research Society (DFG-Forschungsstipedium SI 2502/1-1).

\section*{References}

\bibliography{mybibfile}

\end{document}